\documentclass[journal]{IEEEtran}
\usepackage[small]{caption}
\usepackage{amsmath, amssymb, graphicx, color, cite, subcaption,multicol,wrapfig,amsthm}
\usepackage{algorithm, algorithmic}

\let\OLDthebibliography\thebibliography
\renewcommand\thebibliography[1]{
  \OLDthebibliography{#1}
  \setlength{\parskip}{0pt}
  \setlength{\itemsep}{0pt plus 0.3ex}
  \footnotesize
}

\title{ Deployment and Trajectory Optimization of UAVs: A Quantization Theory Approach}
\author{Erdem Koyuncu,  Maryam Shabanighazikelayeh, and Hulya Seferoglu
    \thanks {The authors are with the   Department of Electrical and Computer Engineering, University of Illinois at Chicago. Emails: \{ekoyuncu, mshaba7, hulya\}@uic.edu. This work has been presented in part at IEEE WCNC in April 2018 \cite{confversion}. This work was supported in part by the NSF Award CCF-1814717.
    } }
 \allowdisplaybreaks

\usepackage{amsmath,amssymb,amsthm}
\newtheorem{proposition}{Proposition}
\newtheorem{theorem}{Theorem}
\newtheorem{example}{Example}
\newtheorem{lemma}{Lemma}

\newcommand{\movement}{M}

\definecolor{superlightgray}{rgb}{.9,0.9,0.9}

\begin{document}

\maketitle
\begin{abstract}
 Optimal deployment and movement of multiple unmanned aerial vehicles (UAVs) is studied. The considered scenario consists of several ground terminals (GTs) communicating with the UAVs using variable transmission power and fixed data rate. First, the static case of a fixed geographical GT density is analyzed. Using high resolution quantization theory, the corresponding best achievable performance (measured in terms of the average GT transmission power) is determined in the asymptotic regime of a large number of UAVs. Next, the dynamic case where the GT density is allowed to vary periodically through time is considered. For one-dimensional networks, an accurate formula for the total UAV movement that guarantees the best time-averaged performance is determined. In general, the tradeoff between the total UAV movement and the achievable performance is obtained through a Lagrangian approach. A corresponding trajectory optimization algorithm is introduced and shown to guarantee a convergent Lagrangian. Numerical simulations confirm the analytical findings. Extensions to different system models  and performance measures are also discussed. 
 \end{abstract}

\section{Introduction}

Unmanned aerial vehicles (UAVs) can be effectively utilized in a variety of wireless communication scenarios. Example applications include providing coverage to geographical areas lacking a wireless infrastructure, relaying to overcome terrain obstacles such as mountains, improving cell edge performance by creating femtocells, among many others \cite{tutopap, yanikomertut, recentuavtut}. 

One of the most distinguishing features of UAV networks is the opportunity of very fast dynamic adaptation to the ever-changing environment through relocation. Environmental variations in this context may include ground terminal (GT) location/density variations, UAV node failures, etc. Although the ability of relocation potentially offers significant performance gains, including improved coverage and rate for GTs, it also comes with many theoretical and practical challenges. Even in a static scenario where the locations or the density of GTs are known and fixed, finding the optimal UAV locations is a non-convex optimization problem whose dimensionality grows with the number of UAVs\cite{tutopap}. Dynamic scenarios further involve optimization of UAV trajectories, thus leading to much more complicated infinite-dimensional optimization problems.

Several approaches to resolve the challenges of UAV deployment/relocation have been proposed. In the case of static deployment, \cite{bor2016efficient, kalantari2016number, lyu1, newref2} consider the optimal placement of UAVs to maximize coverage and propose several algorithms. These works assume that a UAV can cover a GT provided that they are separated no more than a certain distance. In \cite{transporttheory}, the authors consider instead the average throughput as the objective function, and incorporate possible hover time constraints of UAVs into the problem formulation. Static placement of UAVs as relays for offloading cellular traffic \cite{rohde}, or generic multihop communication \cite{koyuncuuavrelay} have also been studied. Random deployments of UAVs are analyzed in \cite{poissonuav} using tools from stochastic geometry. In \cite{uavcache, uavcache2}, the authors consider the optimal deployment of cache-enabled UAVs.

There are also numerous works on dynamic deployment of UAVs. One well-studied scenario is to view UAVs as mobile access points serving GTs. For such a use case, algorithms for UAV coverage under variable coverage radii and possible UAV losses are proposed in \cite{marier1}. For one UAV and one GT, \cite{zeng2017energy} optimizes the UAV trajectory to achieve high throughput with low UAV energy consumption. In \cite{devicetodevice}, the authors consider a single UAV serving a device-to-device communication network. A genetic algorithm for UAV trajectory optimization has been proposed in \cite{anazawa2015trajectory} with the specific goal of restoring network service after natural disasters. In \cite{jiang2012optimization}, the authors optimize the trajectory of a single UAV serving multiple mobile GTs via space-division multiple access. A Kalman filter predicts the future GT locations, which, in turn, determine the UAV trajectory. Given several sensors on a one-dimensional space and one UAV, \cite{aviationtime} determines the time-varying UAV speed that minimizes the data collection time. A related problem is to optimize the UAV trajectories subject to speed constraints \cite{wanguavspeed, uavcomp}. In \cite{completiontime}, the authors consider a UAV multicasting network-coded information to several GTs and the corresponding trajectory optimization problem. An algorithm to minimize the energy consumption of moving the UAVs from one deployment to another can be found in \cite{mozaffari3}.

Several other works have considered the dynamic deployment problem in the context of UAVs serving as communication relays. In particular, for a single source-destination pair and one UAV, \cite{zeng2016throughput} develops a mobile UAV relaying method. The goal is throughput maximization via jointly optimizing UAV trajectory and temporal power allocation. The utilization of UAVs as relays between GTs and a central base station has been studied in \cite{zhan2011wireless}, and a joint heading and adaptive handoff algorithm is proposed. In \cite{athina, shuhang1}, the authors design trajectory optimization algorithms for amplify-and-forward UAVs. For the case of one UAV and a circular trajectory, \cite{dhchoi1} optimizes the speed and load factor of the UAV for maximum energy efficiency. UAVs can also offer computation offloading opportunities in the context of edge computing. A corresponding trajectory optimization problem has been studied in \cite{uavcloud} for the special case of a single UAV.

Despite many recent studies on UAV deployment and trajectory optimization, some of which have been described above, there are many fundamental open problems that are yet to be resolved. In particular, for static networks, there is no general analytical framework that can provide the optimal UAV positions for a given number of UAVs and spatial user density. Also, for the dynamic scenario, an analytical characterization of achievable performance gains are largely missing, and most of the above work relies on numerical methods for optimizing UAV trajectories and determining the resulting performance. Moreover, some of these trajectory optimization algorithms, including \cite{jiang2012optimization, zeng2017energy, zeng2016throughput, shuhang1, dhchoi1, uavcloud}, work only for one UAV, or one dimension \cite{aviationtime}. Some, including \cite{wanguavspeed, uavcomp, zhan2011wireless, mozaffari3, athina, completiontime, newrefzhang}, only consider UAV speed or instantaneous movement/energy limitations, and thus, do not incorporate a constraint on the long-term cost of mobility. Some other algorithms rely on methods such as simulated annealing \cite{rohde} or genetic algorithms \cite{anazawa2015trajectory}, and do not offer convergence guarantees or local optimality.

Quantization theory of data compression and source coding \cite{quantization} has proved to be a very successful analytical tool in addressing many existing problems that involve geographical deployment of agents \cite{VD}; example applications of the theory include the deployment of antenna arrays\cite{koyuncu0}, sensors\cite{skdjlakdjssss}, or general heterogeneous nodes\cite{koyuncu1}. Other applications outside node deployment include image and video compression, classification, and clustering \cite{vqbook}. The preliminaries of the theory as well as other applications can also be found in \cite{vqbook}. The main contribution of this paper is to show that many of the aforementioned open problems on UAV networks can as well be formulated and ultimately resolved using a quantization theory approach. Our setup consists of a density of GTs that are served by an arbitrary number of UAVs. Each GT employs variable-power fixed-rate transmission to ensure outage-free reception at its closest UAV. In the static case, our goal is to find the optimal UAV deployment that minimizes the average GT transmission power, while in the dynamic case, we wish to minimize the time-averaged power consumption subject to a constraint on the \emph{total movement} of the UAVs. The specific main contributions of this paper are then summarized as follows:
\begin{itemize}
\item In the static case, and a uniform distribution of GTs on a line segment on the ground, we determine the optimal deployment of UAVs and the resulting average GT power consumption. For a general one or two dimensional area on the ground and an arbitrary GT distribution, we determine the optimal deployment and the corresponding performance in the asymptotic regime of a large number of UAVs. 
\item In the dynamic case, for any dimension and any time-varying GT densities, we analytically characterize the optimal UAV deployments and the corresponding GT power consumptions in the two extremal cases of no UAV movement and unlimited UAV movement. In the special case of one dimension, our approach leads to an analytical formula for the total UAV movement that guarantees the lowest possible average GT power consumption. In general, compared with the existing studies, the main advantage of our quantization theory approach is that it allows us to obtain similar exact analytical results on the network performance.
\item In order to address moderate UAV movements, we introduce a trajectory optimization algorithm that relies on time discretization, and alternating optimization over each discretized time instance. The algorithm is a descent over the Lagrangian combination of GT power consumption and UAV movement. Optimization at each time instance is carried out through a generalization of Lloyd algorithm \cite{loytoriginal,lbg} in vector quantization. Certain sub-optimization problems that arise in this context are solved in closed form. Our algorithm can numerically provide the optimal UAV trajectories for any dimension and any GT density.
\end{itemize}

Our algorithmic approach is thus aligned with the Voronoi-based coverage control algorithms that were originally envisioned for mobile sensor networks \cite{voro1, voro2, voro3}. These existing studies, however, do not consider total movement constraints and are thus not applicable. Except for the analytical characterization of trajectories and the corresponding total movement for dynamic deployment, where we focus on the case of one dimension, our analysis and algorithms hold for any dimension and any GT density. Also, \cite{bor2016efficient, kalantari2016number, lyu1} only consider static placement of UAVs without any movement, and provide no analytical results on where to place the UAVs and the resulting network performance. In contrast, we consider a dynamic network with UAV mobility and analyze the optimal UAV trajectories and the resulting performance. Also, in the current work, we focus on optimizing the locations and the trajectories of the UAVs. However, in \cite{newref2}, the locations of the UAVs are not optimization variables, but rather are given fixed parameters.

Part of this work has been presented in a conference \cite{confversion}. Compared to \cite{confversion}, most aspects of the trajectory optimization algorithm and the solutions to the accompanying sub-optimization problems are new. In particular, we use a Lloyd algorithm based approach that favors distributed implementation as opposed to the gradient descent based approach in \cite{confversion}. We also provide the implementation details and complexity analysis of the algorithm. Moreover, we extend our results to channel models that incorporate fading, interference, and probabilistic line of sight. We also provide more connections to the existing literature and more detailed numerical simulations.

The rest of this paper is organized as follows: In Section \ref{secsysmodel}, we introduce the system model. Static deployment of UAVs is analyzed in Section \ref{mainseksekstaticdeployment}. We consider the extremal cases of dynamic deployment in Section \ref{mainseksekdynamicdeployment}. Our trajectory optimization algorithm is introduced in Section \ref{secalgogen} to address moderate movement constraints. We provide numerical simulation results in Section \ref{secnumerical}. In Section \ref{secextensions}, we extend our results to different system models. Finally, in Section \ref{secconclusions}, we draw our main conclusions. Some of the technical proofs are provided in the appendices.

\section{System Model}
\label{secsysmodel}
We consider several GTs at zero altitude and several UAVs at a fixed altitude $h\! >\! 0$. Mathematically, the GTs are located on $\mathbb{R}^d,\,d\in\{1,2\}$. While typically one is interested in the case $d=2$, i.e., when the GTs are in general positions on the ground, the case $d=1$ is also relevant: The GTs may be constrained to lie on a line on the ground, e.g., as cars on a straight highway.

We distinguish between what we refer to as the static and the dynamic deployment  scenarios. In static deployment, we assume that the GTs are located on $\mathbb{R}^d$ according to a certain fixed (time-invariant) density function $f$, where $\int_{\mathbb{R}^d} f(q)\mathrm{d}q = 1$. In the more complicated dynamic deployment scenario, we will allow the user density to vary over time. 

\subsection{Static Deployment}
\label{secstaticdeployment}
\begin{figure}[h]
\begin{center}
\scalebox{0.5}{\includegraphics{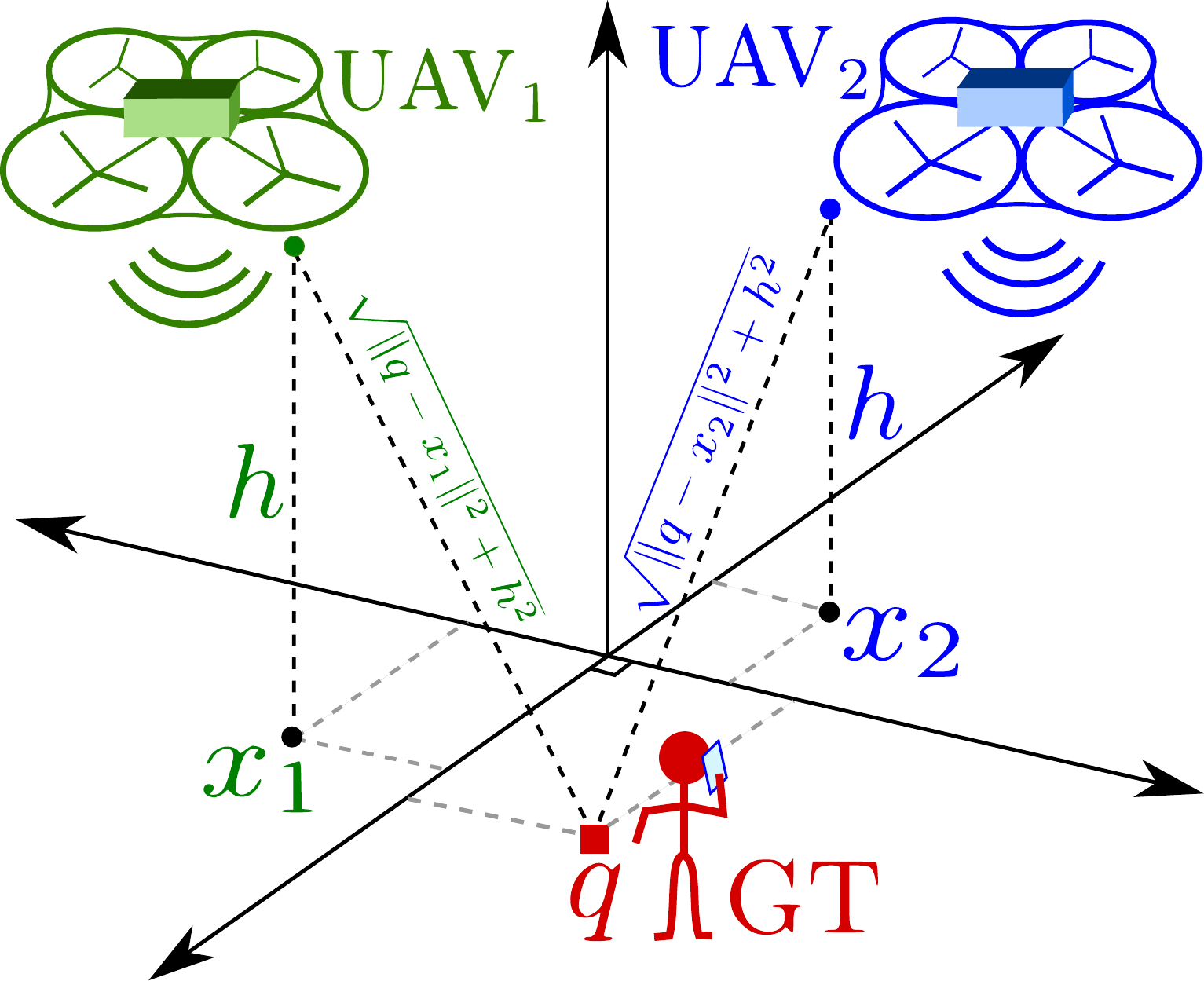}}
\end{center}
\caption{A network of two UAVs serving a GT.}
\label{uavfiggg}
\end{figure}

In order to formally describe the static deployment scenario, let $x_1,\ldots,x_n \in \mathbb{R}^d$ denote the UAVs' (projected) locations on the GT space, measured in meters. The squared Euclidean distance between a GT at $q$ and the $i$th UAV at $x_i$ is then given by $\|x_i - q\|^2 + h^2$. Fig. \ref{uavfiggg} provides an illustration for the special case of $n=2$ and $d=2$. Also, for $d=1$, one can imagine that the horizontal and the vertical positions of UAV $i$ are given by $x_i$ and $h$, respectively.

We consider optimization over only the ground coordinates $x_1,\ldots,x_n$ of the UAVs, while the altitude $h$ of the UAVs is kept fixed. This is because, for all the scenarios that we consider (including the interference-aware model in Section \ref{interferencesec}), decreasing the altitude of any given UAV also decreases the GT's access distance to the UAV, resulting in a better overall  network performance. Therefore, all UAVs should ideally be located on the lowest possible altitude. From this viewpoint, $h$ could also be interpreted as a common minimum altitude constraint that is imposed due to physical obstacles or governmental regulations on the area of interest.

We first consider fixed-rate variable-power transmission at GTs. The case of fixed-power variable-rate transmission will be discussed later. Suppose that a GT at location $q$ wishes to communicate an information-bearing symbol $s$ with rate $\rho$ bits/s/Hz, and transmits with power $P$ W. Due to the aerial nature of the communication system, we assume that there is line of sight between the GT and the UAVs. We also assume that there are no full or even partial physical obstructions between the GTs and the UAVs so that non line of sight effects such as fading are negligible (Extensions to fading or probabilistic line of sight models will be discussed later). The channel input-output relationships are $y_i = s (\|x_i - q\|^2 + h^2)^{-\frac{r}{2}} \sqrt{P} + \eta_i,\,i=1,\ldots,n$, where $y_i$ is the received signal at the $i$th UAV, $r$ is the path loss  exponent, and $\eta \sim \mathcal{CN}(0,1)$ is the noise at UAV $i$. The received signal power at UAV $i$ is thus given by $(\|x_i - q\|^2 + h^2)^{-\frac{r}{2}}P$. Since the noise power is normalized to unity, the signal-to-noise ratio (SNR) at UAV $i$ also equals $(\|x_i - q\|^2 + h^2)^{-\frac{r}{2}}P$. Reliable communication between the GT and the UAV is possible provided that the channel capacity between the GT and the UAV is at least $\rho$, or, mathematically if 
\begin{align}
\label{capacitycriterion}
\log_2(1 + (\|x_i - q\|^2 + h^2)^{-\frac{r}{2}}P) \geq \rho. 
\end{align}

According to (\ref{capacitycriterion}), for reliable communication to the UAV at $x_i$, the transmission power of the GT  should satisfy $P \geq (2^{\rho} - 1)(\|x_i - q\|^2 + h^2)^{\frac{r}{2}}$. The minimum transmission power that guarantees successful data reception at one or more of the UAVs is therefore $\min_i (2^{\rho} - 1)(\|x_i - q\|^2 + h^2)^{\frac{r}{2}}$. Averaging out the GT density, and setting $\rho = 1$ throughout the paper without loss of generality, the average transmission power of GTs given UAV locations $\mathbf{x} \triangleq [x_1 \cdots x_n]$ and density $f$ is 
\begin{align}
\label{avpow}
P(\mathbf{x}, f) \triangleq \int_{\mathbb{R}^d} \min_i (\|x_i - q\|^2 + h^2)^{\frac{r}{2}} f(q)\mathrm{d}q.
\end{align}
The static deployment problem is then to find the optimal UAV locations that minimize the average GT power consumption. In other words, we wish to determine $P^{\star}(f) \triangleq \min_{\mathbf{x}}P(\mathbf{x}, f)$, and the optimal deployments $\mathbf{x}^{\star}$ that achieve $P(\mathbf{x}^{\star}, f)  = P^{\star}(f)$.

\subsection{Dynamic Deployment}
\label{secdynamicdeployment}
In practice, the GT density may vary over time. For example, in daily urban communications, the GT density over highways will be higher during rush hours, when compared to nighttime. In order to model such scenarios, we let $f_t$ denote the GT density function at time $t$. We assume $f_t$ is periodic over a time interval of length $T$, i.e., $f_t(q) = f_{t+T}(q),\,\forall t,\,\forall q$. For example, one may set $T = 24$ hours for the urban highway communication scenario, as the traffic or GT density at a given highway stretch can be assumed to be the same for the same hours of different days.  In general, we assume that $f_t(q)$ is continuous in both $t$ and $q$. Thus, the GT density does not experience abrupt changes over space or time. In practice, for a fixed number of GTs, variations of the GT density correspond to GT mobility.

Let $x_{t,i}$ denote the location of UAV $i$ at time $t$, and $\mathbf{x}_t = [x_{t,1} \cdots x_{t,n}]$ denote the vector of UAV locations at time $t$. 
The power consumption of GTs at time $t$ is $P(\mathbf{x}_t, f_t)$. The average power consumption over time can be expressed as
\begin{align}
\label{powconsumpt}
Q \triangleq \frac{1}{T} \int_0^T P(\mathbf{x}_t, f_t) \mathrm{d}t.
\end{align}

Throughout the paper, we assume that the start and the end locations of each UAV over one period is the same; i.e., $x_{0,i} = x_{T,i},\,\forall i$. The time-averaged distance traversed by the $i$th UAV can then be calculated to be the line integral
\begin{align}
\label{uavimovement}
\movement_i \triangleq \frac{1}{T} \int_0^T \sqrt{\sum_{j=1}^d \left|\frac{\partial x_{t,i,j}}{\partial t}\right|^2 } \mathrm{d}t,
\end{align}
where $x_{t,i,j}$ represents the $j$th component of $x_{t,i}$. In this case, the goal is to find the optimal UAV trajectories that minimize the average GT power consumption $Q$ subject to a constraint  
$\sum_{i=1}^n \movement_i \leq M$ on the total UAV movement $\sum_{i=1}^n \movement_i$, where $M \geq 0$ is given. Note that, for the special case of no UAV movement $M=0$, and a time-invariant density $f_t = f,\,\forall t$, the dynamic deployment scenario reduces to the static deployment scenario described in Section \ref{secstaticdeployment}.
\section{Optimization of a Static Deployment}
\label{mainseksekstaticdeployment}
We begin with the simpler scenario of a static deployment and its optimization. Namely, we study the minimization of (\ref{avpow}) with respect to UAV locations $\mathbf{x}$. We begin by considering the degenerate case $h=0$, in which case we can imagine that the network consists of unmanned ground vehicles (UGVs) instead of UAVs. The analysis of such a UGV scenario will be very useful for our analysis of the UAV case $h > 0$. Later, we will also apply the results of this section to the dynamic deployment problem for UGVs in Sections \ref{mainseksekdynamicdeployment} and \ref{secalgogen}. In this context, although there are some existing works on static UGV deployment (see \cite{koyuncu1} and the references therein), none consider our dynamic deployment scenario with a total movement constraint.

\subsection{The UGV Case $h=0$}
\label{secdegenerate}
For $h=0$, the cost function in (\ref{avpow}) becomes $P(\mathbf{x}, f) = \int_{\mathbb{R}^d} \min_i \|x_i - q\|^r f(q)\mathrm{d}q$. This expression is the well-known average $r$th power distortion of a quantizer whose reproduction points are $x_1,\ldots,x_n$ for a given source density $f$. Finding its exact minimizers and the corresponding minimum distortions is possible only for a few special cases. In particular, if $f(q) =  \mathbf{1}(q\in[0,1])$ is the one-dimensional uniform density, then the optimal reproduction points are given by the uniform quantizer codebook $\mathbf{x}_u = [\frac{1}{2n}\,\frac{3}{2n}\,\cdots\,\frac{2n-1}{2n}]$ with $P(\mathbf{x}_u,f) = \frac{1}{(1+r)(2n)^r}$. 

For a general uniform density, we have the following result of Bennett \cite{bennett1} and Zador {\cite{zador1}}. Given $A \subset \mathbb{R}^d$, let $m(A) \triangleq \int_A \|q\|^r \mathrm{d}q/(\int_A \mathrm{d}q)^{\frac{d+r}{d}}$ denote the normalized $r$th moment of $A$.
\begin{proposition}
\label{piqjwpeojqw}
Let $f(q) =  \mathbf{1}(q\in[0,1]^d)$, $h=0$. As $n\rightarrow\infty$,  we have $P^{\star}(f) = \kappa_{rd} n^{-\frac{r}{d}} + o(n^{-\frac{r}{d}})$, where $\kappa_{rd}$ depends only on $r$ and $d$. In particular, $\kappa_{r1} =  \frac{2^{-r}}{1+r}$ and $\kappa_{r2}$ are the normalized moments of the origin-centered interval and the origin-centered regular hexagon, respectively.
\end{proposition}
This implies that for $d=2$ and a uniform distribution, the best arrangement of quantization points is asymptotically the regular hexagonal lattice. Equivalently, for a two-dimensional uniform GT density, the best arrangement of UAVs is asymptotically the regular hexagonal lattice. 

Making the transition from uniform to non-uniform $f$ can be accomplished using the idea of point density functions. In detail, one assumes the existence of a function $\lambda(q)$ such that the cube $[q,q+\mathrm{d}q]$ of volume $\mathrm{d}q$ contains $n\lambda(q)\mathrm{d}q$ reproduction points with $\int_{\mathbb{R}^d} \lambda(q)\mathrm{d}q = 1$. Since $f$ should be approximately uniform on $[q,q+\mathrm{d}q]$, the conditional average distortion on $[q, q+\mathrm{d}q]$ is $\kappa_{rd} (n \lambda(q))^{-\frac{r}{d}} + o(n^{-\frac{r}{d}})$ by Proposition \ref{piqjwpeojqw}. Averaging out the density, we obtain the formula 
\begin{align}
 \widetilde{P}(\lambda) \triangleq \kappa_{rd}n^{-\frac{r}{d}}\int_{\mathbb{R}^d} f(q) \lambda^{-\frac{r}{d}}(q)\mathrm{d}q + o(n^{-\frac{r}{d}}).
\end{align}
for the average distortion given $\lambda$. Using reverse H\"{o}lder's inequality, we have
\begin{align}
\label{gendistodim}
\widetilde{P}(\lambda) \geq \kappa_{rd} n^{-\frac{r}{d}} \|f\|_{\frac{d}{d+r}}+ o(n^{-\frac{r}{d}}),
\end{align}
where $\|f\|_{\alpha} \triangleq (\int_{\mathbb{R}^d} (f(q))^{\alpha} \mathrm{d}q)^{\frac{1}{\alpha}}$ is the $\alpha$-norm of the density $f$. In (\ref{gendistodim}), equality holds if
\begin{align}
\label{optpdf}
\lambda(q) = f^{\frac{d}{d+r}}(q)/\textstyle \int_{\mathbb{R}^d} f^{\frac{d}{d+r}}(q') \mathrm{d}q',\,\forall q
\end{align}
Thus, the minimum distortion $\kappa_{rd} n^{-\frac{r}{d}} \|f\|_{\frac{d}{d+r}} + o(n^{-\frac{r}{d}})$ is achieved by the point density  in (\ref{optpdf}).
\subsection{The UAV Case $h>0$}
We now consider the case $h > 0$. We begin with the simple case of a uniform one-dimensional distribution.  The proof of the following proposition can be found in Appendix A. 
\begin{proposition}
\label{onedimuniprop}
Let $f(q) =  \mathbf{1}(q\in[0,1])$. A minimizer of (\ref{avpow}) is the uniform quantizer codebook $\mathbf{x}_u$. In other words, the optimal placement for $n$ UAVs and uniform GT density is given by $\mathbf{x}_u$. The corresponding minimum average power is  $P(\mathbf{x}_u,f) = 2n \int_{0}^{\frac{1}{2n}} (u^2 + h^2)^{\frac{r}{2}} \mathrm{d}u$. 
\end{proposition}

For a general $d$ and $f$, we observe that if $x_1,\ldots,x_n$ is an optimal deployment, the set of points $q$ with the property that $\min_i \|q - x_i\| \rightarrow 0$ as $n\rightarrow\infty$ has probability one (Otherwise, there is a constant $a > 0$ such that with some positive probability $\epsilon > 0$, one has $\min_i \|q - x_i\| > a$ infinitely often. This implies $P^{\star}(f) > \epsilon a$ infinitely often, contradicting Proposition \ref{piqjwpeojqw}). This amounts to the intuitive observation that every GT should be assigned to a closer UAV as the number of available UAVs grows to infinity. As a result, we may use the Taylor series expansion 
\begin{multline}
\left(\min_i \|x_i - q\|^2 + h^2\right)^{\frac{r}{2}} = h^r + \tfrac{1}{2}rh^{r-2}\min_i \|x_i - q\|^2 + \\ o\left(\min_i \|x_i - q\|^2\right)
\end{multline}
so that, substituting to (\ref{avpow}), we have
\begin{multline}
\label{avpowwewew}
P(\mathbf{x}, f) =  h^r + \tfrac{1}{2}rh^{r-2} \int_{\mathbb{R}^d} \min_i \|x_i - q\|^2 f(q)\mathrm{d}q + \\ \int_{\mathbb{R}^d} o\left(\min_i \|x_i - q\|^2\right) f(q)\mathrm{d}q.
\end{multline}
Using (\ref{gendistodim}) and (\ref{optpdf}), we can then obtain the following theorem.
\begin{theorem}
\label{t1}
As $n\rightarrow\infty$, we have 
\begin{align}
\label{asymptoticdistoformul}
P^{\star}(f) \!=\! \left\{ \begin{array}{rl} \!\!\!
\kappa_{rd}n^{-\frac{r}{d}}\|f\|_{\frac{d}{d+r}} \!+\! o(n^{-\frac{r}{d}}),\!\!\! & h \!=\! 0, \\ \!\!h^r \!+\! \frac{rh^{r-2}\kappa_{2d}}{2}n^{-\frac{2}{d}}\|f\|_{\frac{d}{d+2}} \!+\! o(n^{-\frac{2}{d}}),\!\!\! & h \!>\! 0. \\\end{array}\right.\!\!\!
\end{align}
The optimal point (UAV) density function is given by 
\begin{align}
\label{lambdastarqf}
\lambda^{\star}(q;f) \triangleq \left\{ \begin{array}{rl} 
f^{\frac{d}{d+r}}(q) / \int_{\mathbb{R}^d} f^{\frac{d}{d+r}}(q') \mathrm{d}q', & h = 0, \\ f^{\frac{d}{d+2}}(q) / \int_{\mathbb{R}^d} f^{\frac{d}{d+2}}(q') \mathrm{d}q', & h > 0. \\\end{array}\right.
\end{align}
Hence, as $n\rightarrow\infty$, for any $q$, the infinitesimal $[q,q+\mathrm{d}q]$ should contain $n\lambda^{\star}(q;f)\mathrm{d}q$ UAVs.
 \end{theorem}
This provides a complete asymptotic characterization of the achievable GT power consumption and the corresponding optimal UAV configuration.
\section{Optimization of a Dynamic Deployment: Extremal Cases}
\label{mainseksekdynamicdeployment}
We now consider the dynamic scenario, where the GT density varies periodically over time. As discussed in Section \ref{secdynamicdeployment}, the goal in this case is to minimize the time-averaged power consumption $Q$ in (\ref{powconsumpt}), subject to the constraint  $\sum_{i=1}^n \movement_i \leq M$ on the total movement  of UAVs. Here, $\movement_i$ denotes the total movement of the $i$th UAV, and has been defined in (\ref{uavimovement}). Given $\movement \geq 0$, we use the notation $Q^{\star}(\movement)$ to denote the minimum of (\ref{powconsumpt}) subject to $\sum_{i=1}^n \movement_i \leq \movement$.  In particular, in this section, we consider the two extremal cases $\movement = 0$ and $\movement  \rightarrow \infty$. The remaining moderate cases will be discussed in Section \ref{secalgogen}. 
\subsection{No UAV movement: $\movement = 0$}
The case $\movement = 0$ corresponds to a scenario where we do not allow any UAV movement. Equivalently, the UAV locations are fixed over time as $\mathbf{x}_t = \mathbf{x}',\forall t$ for a collection $\mathbf{x}' = [x_{1}'\cdots x_{n}']$ of UAV locations to be optimized. By (\ref{avpow}) and (\ref{powconsumpt}), we  have
\begin{align}
\label{qpeuqpwoeuqoweqweqwe}
Q^{\star}(0) \!= \!\min_{\mathbf{x}'} \frac{1}{T} \!\!\int_0^T \!\!\!\int_{\mathbb{R}^d}\! \min_i (\|x_i' \!-\! q\|^2 \!+\! h^2)^{\frac{r}{2}} f_t(q)\mathrm{d}q \mathrm{d}t.
\end{align}
Now, let $\overline{f}(q) \triangleq \frac{1}{T}\int_0^T f_t(q)\mathrm{d}t$ be the ``time-averaged density.'' Note that $\int_{\mathbb{R}^d}\overline{f}(q)\mathrm{d}q = 1$ so that $\overline{f}$ is a valid density function. According to (\ref{avpow}) and (\ref{qpeuqpwoeuqoweqweqwe}), we have
$Q^{\star}(0) = \min_{\mathbf{x}'} P(\mathbf{x}',\overline{f})$, and optimizing over $\mathbf{x}'$ leads to the following.
\begin{proposition}
\label{qpowennndndndn1}
The optimal GT power consumption without any UAV movement is $Q^{\star}(0) \!=\! P^{\star}(\overline{f})$.
\end{proposition}
Theorem \ref{t1} can be applied to provide an asymptotically tight expression for $ P^{\star}(\overline{f})$.
\begin{example}
\label{example1}
Let us consider a one-dimensional network $d=1$, a period of length $T=2$ with path loss exponent $r=2$.  For a simpler exposition, we further consider an UGV network where $h=0$. Let the time-varying GT density be given by
$f_t(q) = (1+3|t|)(q-2+2|t|)^{3|t|}, \,q\in[2-2|t|,3-2|t|],t\in[-1,1]$. This defines shifted power-law densities. For example for $t=-1$, we obtain the density $f_1(q) = 4q^3,\,q\in[0,1]$, and for $t=0$, we obtain $f_0(q) = 1,\,q\in[2,3]$. The time-averaged density $\overline{f}$ as well as its $\frac{1}{3}$-norm $\|\overline{f}\|_{\frac{1}{3}} = (\int_{\mathbb{R}} (f(q))^{\frac{1}{3}} \mathrm{d}q)^{3} \approx 6.08$ can be found by numerical integration. By Proposition \ref{qpowennndndndn1} and Theorem \ref{t1}, it follows that 
\begin{align}
\label{bestpowinonedimzeromvmt}
Q^{\star}(0) \approx \frac{6.08}{12} \frac{1}{n^2} + o\left(\frac{1}{n^2}\right) = \frac{0.507}{n^2} + o\left(\frac{1}{n^2}\right). 
\end{align}
The optimal point (UAV) density function is $\lambda^{\star}(q,\overline{f})$, as defined in (\ref{lambdastarqf}). Different GT densities can be analyzed in the same manner by using the formulae in Proposition \ref{qpowennndndndn1} and Theorem \ref{t1}. \qed
\end{example}
\subsection{Unlimited UAV movement: $\movement \rightarrow \infty$}
We now allow an unlimited UAV movement to obtain the minimum possible time-averaged GT power consumption. For this purpose, at each time $t$, we use the UAV locations that provide the minimum ``instantaneous'' GT power consumption. This results in the time-averaged power
\begin{align}
\label{oiqwe222oqwiheqw}
Q^{\star}(\infty) = \frac{1}{T} \int_0^T P^{\star} (f_t) \mathrm{d}t.
\end{align}
We recall that Theorem \ref{t1} provides an asymptotic expression for the integrand  $P^{\star} (f_t)$. This can be substituted to (\ref{oiqwe222oqwiheqw}) for an asymptotically tight characterization of  $Q^{\star}(\infty)$.

We now argue that (\ref{oiqwe222oqwiheqw}) is, in fact, achievable with a finite amount of total movement as well. In other words, there is a constant $\overline{\movement} > 0$ such that $Q^{\star}(\movement) = Q^{\star}(\infty)$ for every $\movement \geq \overline{\movement}$. The idea is to observe that the GT density $f_t$ at time $t$ is not ``vastly different'' than the GT density $f_{t+\mathrm{d}t}$ at time $t+\mathrm{d}t$. This stems from our practical assumption in Section \ref{secdynamicdeployment} that the spatiotemporal density $f_t(q)$ is continuous in both space and time. As a result, we expect the optimal location for each UAV to be a well-behaved continuous function of time, resulting in a finite amount of total UAV movement.

We utilize high-resolution quantization theory to estimate $\overline{\movement}$. The key is to recover the location of each UAV at any given point $t$ in time through the optimal quantizer point density function at time $t$. Namely, let $\mathbf{x}_t^{\star} \triangleq [x_{t,1}^{\star}\cdots x_{t,n}^{\star}]$ denote the optimal UAV locations at time $t$ for density $f_t$. We first consider the case of one dimension $d=1$.  Without loss of generality, suppose $x_{t,1}^{\star} \leq \cdots \leq x_{t,n}^{\star}$. Given $x\in[0,1]$, let
$\Lambda_{\mathrm{inv}}^{\star}(x;f_t)$ be the unique real number that satisfies
\begin{align}
\label{lksjdalksjdasdasd}
\int_0^{\Lambda_{\mathrm{inv}}^{\star}(x;f_t)} \lambda^{\star}(q;f_t) \mathrm{d}q = x,
\end{align}
where $\lambda^{\star}(q;f_t)$ is the optimal point density function for $f_t$, as defined in (\ref{lambdastarqf}) of Theorem \ref{t1}. Note that $\Lambda_{\mathrm{inv}}^{\star}(x;f_t)$ is the inverse of the cumulative distribution function $u\rightarrow \int_0^{u} \lambda^{\star}(q;f_t) \mathrm{d}q$. Our idea is to approximate the optimal UAV locations via
\begin{align}
\label{ldkajslkdjaklsdj1}
x_{t,i}^{\star} \simeq \widetilde{x}_{t,i} \triangleq \Lambda_{\mathrm{inv}}^{\star}\left(\frac{2i-1}{2n};f_t\right),\,i=1,\ldots,n.
\end{align}
Note that if $U$ is a random variable that is uniformly distributed on $[0,1]$, then, according to the inverse transform sampling method, the random variable $\Lambda_{\mathrm{inv}}^{\star}(U;f_t)$ is distributed according to the density function $\lambda^{\star}(q,f_t)$. The transformation in (\ref{ldkajslkdjaklsdj1}) can thus be considered to be a ``deterministic version'' of inverse transform sampling, where the uniform random variable $U$ is replaced with the uniform quantizer with reproduction points $\frac{2i-1}{2n},\,i=1,\ldots,n$. The resulting estimates $\widetilde{x}_{t,1},\ldots,\widetilde{x}_{t,n}$ are consistent with the density function $\lambda^{\star}(q;f_t)$ in the sense that for every $q$ and $\epsilon > 0$, the fraction $\frac{1}{n}|\{i:\widetilde{x}_{t,i} \in (q,q+\epsilon)\}|$ of UAVs that are located on $(q,q+\epsilon)$ converges to  $\lambda^{\star}(q;f_t)\epsilon$ as $n\rightarrow\infty$. Now, substituting (\ref{ldkajslkdjaklsdj1}) to (\ref{uavimovement}), we obtain the following result.
\begin{theorem}
\label{t2}
Let $d=1$. As $n\rightarrow\infty$, the minimum possible average power consumption of $Q^{\star}(\infty) = \frac{1}{T} \int_0^T P^{\star} (f_t)$
is achievable with a total movement of
\begin{align}
\overline{\movement}_i\triangleq \frac{1}{T} \int_0^T  \left| \frac{\partial \Lambda_{\mathrm{inv}}^{\star}\left(\frac{2i-1}{2n};f_t\right)}{\partial t} \right| \mathrm{d}t
\end{align}
for the $i$th UAV. Correspondingly, $\overline{\movement} = \sum_{i=1}^n \overline{\movement}_i$.
\end{theorem}
\begin{example}
\label{example2}
We continue the setup in Example \ref{example1}. We have
\begin{align}
\nonumber \|f_t\|_{\frac{1}{3}} & = (1+3|t|) \left(\int_{2-2|t|}^{3-2|t|} (q-2+2|t|)^{|t|}\mathrm{d}q\right)^3 \\ \nonumber & = (1+3|t|) \left(\int_{0}^{1} q^{|t|}\mathrm{d}q\right)^3 \\ \label{ftonethirdeval}
 & = \frac{1+3|t|}{(1+|t|)^3}.
\end{align}
This yields
\begin{align}
\nonumber Q^{\star}(\infty) & = \frac{1}{2} \int_{-1}^1 P^{\star} (f_t) \mathrm{d}t \\  & = \frac{1}{24n^2} \int_{-1}^1 \|f_t\|_{\frac{1}{3}} \mathrm{d}t + o\left(\frac{1}{n^2}\right) \\ \nonumber & =  \frac{1}{24n^2} \int_{-1}^1 \frac{1+3|t|}{(1+|t|)^3} \mathrm{d}t + o\left(\frac{1}{n^2}\right) \\ \label{bestpowinonediminfmvmt}  & = \frac{1}{16n^2} + o\left(\frac{1}{n^2}\right).
\end{align}
The first three equalities follow from (\ref{oiqwe222oqwiheqw}), (\ref{asymptoticdistoformul}), and (\ref{ftonethirdeval}), respectively. In order to estimate $\overline{\movement}$, we use the formula (\ref{lambdastarqf}) to first calculate $\lambda^{\star}(q,f_t) = (1+|t|)(q-2+2|t|)^{|t|}$. In the light of (\ref{lksjdalksjdasdasd}), we then solve for $\theta$ in the integral equality
$\int_0^{\theta} (1+|t|)(q-2+2|t|)^{|t|} \mathrm{d}q = x$ to obtain the inverse cumulative distribution function
$\Lambda_{\mathrm{inv}}^{\star}(x,f_t) = \theta = 2-2|t|+x^{\frac{1}{1+|t|}},\,x\in[0,1],\,t\in[-1,1]$. 
 According to (\ref{ldkajslkdjaklsdj1}), we can then obtain 
 \begin{align}
\nonumber  \widetilde{x}_{t,i} & =\Lambda_{\mathrm{inv}}^{\star}\left(\frac{2i-1}{2n};f_t\right) \\  \label{optuavtrajsjsjss}
 & =  2-2|t|+\left(\frac{2i-1}{2n}\right)^{\frac{1}{1+|t|}},\,i=1,\ldots,n. 
 \end{align}
 Note that, for a fixed index $i$, the function $t \mapsto \widetilde{x}_{t,i}$ is symmetric around the origin and decreases on $[0,1]$. Theorem \ref{t2} combined with the fundamental theorem of calculus then yields
 \begin{align}
  \nonumber \overline{M}_i & = \frac{1}{2}\int_{-1}^1 \left| \frac{\partial \widetilde{x}_{t,i}}{\partial t} \right| \mathrm{d}t \\ \nonumber & =  \widetilde{x}_{0,i} - \widetilde{x}_{1,i} \\ &  = 2 +  \frac{2i-1}{2n}- \left(\frac{2i-1}{2n}\right)^{\frac{1}{2}}.
  \end{align}
  Thus, the power consumption of $Q^{\star}(\infty)$ is achievable with a total UAV movement of 
  \begin{align}
  \overline{M} = \sum_{i=1}^n \overline{M}_i = 2n + \sum_{i=1}^n \left( \frac{2i-1}{2n}- \left(\frac{2i-1}{2n}\right)^{\frac{1}{2}} \right), 
  \end{align}
 and the optimal trajectories are given by (\ref{optuavtrajsjsjss}). As $n\rightarrow\infty$, we have 
 \begin{align}
 \label{totmvmtformula}
 \frac{\overline{M}}{n} \rightarrow 2 + \int_0^1 (x - \sqrt{x})\mathrm{d}x = \frac{11}{6}.  
 \end{align}
 Therefore, for a per-UAV movement of $\frac{11}{6}$, a GT power consumption of roughly $\frac{1}{16n^2}$ is achievable. On the other hand, Example \ref{example1} shows that without any UAV movement, a GT power consumption of roughly $\frac{0.507}{n^2}$ is achievable. For the particular scenario in Examples \ref{example1} and \ref{example2}, allowing mobility of access points thus potentially yields an $8$-fold reduction in the GT power consumption. \qed
\end{example}

The arguments that we have used to obtain Theorem \ref{t2} are not immediately applicable to the case of two dimensions. The main difficulty is to find a simple analogue of (\ref{ldkajslkdjaklsdj1}) that can faithfully extract the optimal UAV locations from the optimal UAV density functions. We leave a resolution of this problem as future work. Nevertheless, $\overline{\movement}$ and $Q^{\star}(\overline{\movement})$ can still be numerically approximated for two dimensional densities as we show in Section \ref{secnumerical}.  
\section{Optimization of a Dynamic Deployment: Moderate Distances}
\label{secalgogen}
We recall that our goal in the dynamic deployment scenario is to find the minimum average GT power consumption $Q^{\star}(M)$ subject to the total movement constraint $M$ on the UAVs. In the previous section, we have analytically characterized the achievable performance in the extremal cases of no UAV movement $M=0$ and unlimited UAV movement $M = \infty$. In particular, we have shown that there exists a sufficient amount of total movement $\overline{M}$ such that $Q^{\star}(\overline{M}) = Q^{\star}(\infty)$. We now consider the achievable performance between the two extremal cases. In other words, we consider the moderate distances regime $0 < M < \overline{M}$.

A precise analytical characterization of the achievable performance appears to be very challenging for the case $0 < M < \overline{M}$. We thus mainly follow a numerical approach. Specifically, we introduce a Lagrangian-based descent algorithm for trajectory optimization. 


\subsection{Outline of an Algorithm for Trajectory Optimization}
\label{secalgo1}
Our general strategy for trajectory optimization is to follow the classical Lagrangian approach of constrained optimization. Namely, we combine the power consumption (objective) function $Q$ in (\ref{powconsumpt}) and the movement (constraint) function $\sum_{i=1}^n M_i$  through the Lagrangian
\begin{multline}
\label{lagranjiyin}
 Q +\ell \sum_{i=1}^n M_i = \\  \frac{1}{T} \int_0^T \!\!\left[P(\mathbf{x}_t, f_t)  + \ell \sum_{i=1}^n \sqrt{\sum_{j=1}^d \left|\frac{\partial x_{t,i,j}}{\partial t}\right|^2 } \,\,\right] \mathrm{d}t.
\end{multline}

Minimizing the Lagrangian for different values of the Lagrange multiplier $\ell >0$ enables travel over the $(M, Q^{\star}(M))$ tradeoff curve: For example, a small $\ell$ does not penalize the total movement as much as a larger $\ell$ does. It thus results in a lower power consumption compared to the case of a larger $\ell$, albeit at the expense of more movement. The formulation in (\ref{lagranjiyin}) thus resembles the Lagrangian formulation of the entropy-constrained quantizer design problem \cite{ecvq}.

Minimizing (\ref{lagranjiyin}) requires optimization over the uncountably many variables $x_{t,i},\,t\in[0,T],\,i\in\{1,\ldots,n\}$, and is thus infeasible. The first step towards a feasible optimization is to discretize the continuous time interval $[0,T]$ to the set of discrete time instances $\{\frac{kT}{K}:k\in\{0,\ldots,K-1\}\}$, where $K \geq 2$ is a natural number. This results in the discrete-time Lagrangian
\begin{align}
\label{disctimelagrange}
\mathcal{L}  \triangleq \frac{1}{K} \sum_{k=0}^{K-1} P(\mathbf{y}_{k}, \hat{f}_k)  + \frac{\ell}{K} \sum_{k=0}^{K-1} \sum_{i=1}^n \|y_{k,i} - y_{k-1,i}\|, 
\end{align}
where the discrete time $k$ corresponds to the continuous time $\frac{kT}{K}$, and the optimization is over $\mathbf{y}_k \triangleq \mathbf{x}_{\frac{kT}{K}} = [y_{k,i}\cdots y_{k,n}]$ with density $\hat{f}_k \triangleq f_{\frac{kT}{K}}$. Also, for a simple notation, we have omitted to indicate that all $k$-dependent indices are evaluated modulo $K$. For example, for $k=0$, the discrete time index $k-1=-1$ is the same as the discrete time index $-1\,\mathrm{mod}\, K = K-1$.

It can be shown that, under some technical conditions on $x_{t,i},\,i=1,\ldots,n$ and $f_t$, such as continuity in $t$, the discrete time Lagrangian converges to the continuous time Lagrangian as the number of time steps $K$ grows to infinity. We thus expect the minimizers of (\ref{disctimelagrange}) and  (\ref{lagranjiyin}) to coincide asymptotically as $K\rightarrow\infty$. In other words, we can obtain the optimal trajectories for the original continuous-time problem formulation in Section \ref{secsysmodel} as $K\rightarrow\infty$. Note that this is different than our approach in the previous sections, where we considered asymptotically large number of UAVs $n\rightarrow\infty$. Still, the direct minimization of (\ref{disctimelagrange}) is a $dnK$ dimensional optimization problem. In order to further reduce the dimensionality, we define
\begin{multline}
\label{ellkey}
 \mathcal{L}_k \triangleq \frac{1}{K} P(\mathbf{y}_{k}, \hat{f}_k)  + \frac{\ell}{K}   \sum_{i=1}^n \|y_{k,i} - y_{k-1,i}\| + \\ \frac{\ell}{K}   \sum_{i=1}^n \|y_{k,i} - y_{k+1,i}\|,k=1,\ldots,K,
\end{multline}
and note that $\mathcal{L}$ depends on $\mathbf{y}_k$ only through $\mathcal{L}_k$. The quantity $\mathcal{L}_k$ can be considered to be the Lagrangian cost at   time instance $k$. Our algorithm is then to perform alternating optimization over the discrete time instances, as shown in Algorithm \ref{algorithm1}. 

\begin{algorithm}
\begin{algorithmic}[1]
\vspace{10pt}\STATE Initialize $\mathbf{y}_0,\ldots,\mathbf{y}_{K-1}$. Set $\mathtt{maxEpochs}$.
\FOR{$\mathtt{epochs} = 1$ \textbf{to} $\mathtt{maxEpochs}$}
\STATE Update the UAV deployments as $\mathbf{y}_k\leftarrow\arg\min_{\mathbf{y}_k} \mathcal{L}_k,\,k=0,\ldots,K-1$.
\ENDFOR
\end{algorithmic}
\caption{Trajectory Optimization}
\label{algorithm1}
\end{algorithm}

In detail, we begin with an initial (e.g., random) guess on trajectories $\mathbf{y}_{0},\ldots,\mathbf{y}_{K-1}$. For the sequence of time indices $k=0,1,\ldots,K\!-\!1,0,1,\ldots,K\!-\!1,\ldots,$ we minimize $\mathcal{L}_k$ over $\mathbf{y}_k$, while keeping all $\mathbf{y}_i,\,i\neq k$ fixed (The specific manner in which we perform the minimization will be discussed later on). Since each step minimizes $\mathcal{L}_k$ over $\mathbf{y}_k$ for some $k\in\{0,\ldots,K-1\}$, and the dependence of $\mathcal{L}$ on $\mathbf{y}_k$ is only through $\mathcal{L}_k$, the process guarantees a non-increasing $\mathcal{L}$. To be more precise, let us write $\mathcal{L}(\mathbf{Y} )$ to signify the dependence of $\mathcal{L}$ in  (\ref{disctimelagrange})  on the UAV trajectories $\mathbf{Y} \triangleq [\mathbf{y}_1 \cdots \mathbf{y}_K]$. Also, given arbitrary initial conditions in Algorithm \ref{algorithm1},  let $\mathbf{y}_{p,k}$ be the vector of UAV locations at discrete time $k$, and let $\mathbf{Y}_p \triangleq [\mathbf{y}_{p,1} \cdots \mathbf{y}_{p,K}]$ denote the UAV trajectories at the end of epoch $p \geq 1$. Then, we have $\mathcal{L}(\mathbf{Y}_{p+1}) \leq \mathcal{L}(\mathbf{Y}_{p})$ for every $p \geq 1$. Since $\mathcal{L}(\mathbf{Y}) \geq 0$ obviously holds for any collection of trajectories $\mathbf{Y}$, it follows by the monotone convergence theorem that the sequence of costs $\mathcal{L}(\mathbf{Y}_1),\mathcal{L}(\mathbf{Y}_2),\ldots$ provided by the algorithm converges.

In our numerical experiments, we have observed that the algorithm also provides convergent trajectories as well; i.e., the sequence $\mathbf{Y}_1,\mathbf{Y}_2,\ldots$ also converges. A formal proof of this observation will remain as an interesting direction for future research. Also, note that in Algorithm \ref{algorithm1}, we call one pass over all discrete time instances as one epoch of optimization. The algorithm terminates after a certain number of epochs that is to be chosen depending on the input parameters. Different termination criteria (such as the convergence of trajectories or the cost $\mathcal{L}$) can also be considered. Obviously, due to the non-convex non-linear nature of the trajectory optimization problem, the resulting trajectory may not necessarily be the globally-optimal solution. Nevertheless, since Algorithm \ref{algorithm1} provides a monotonically non-increasing cost function, it can improve almost any initial UAV trajectories for a better network performance.

\begin{figure*}
\begin{center}
\scalebox{0.7}{\includegraphics{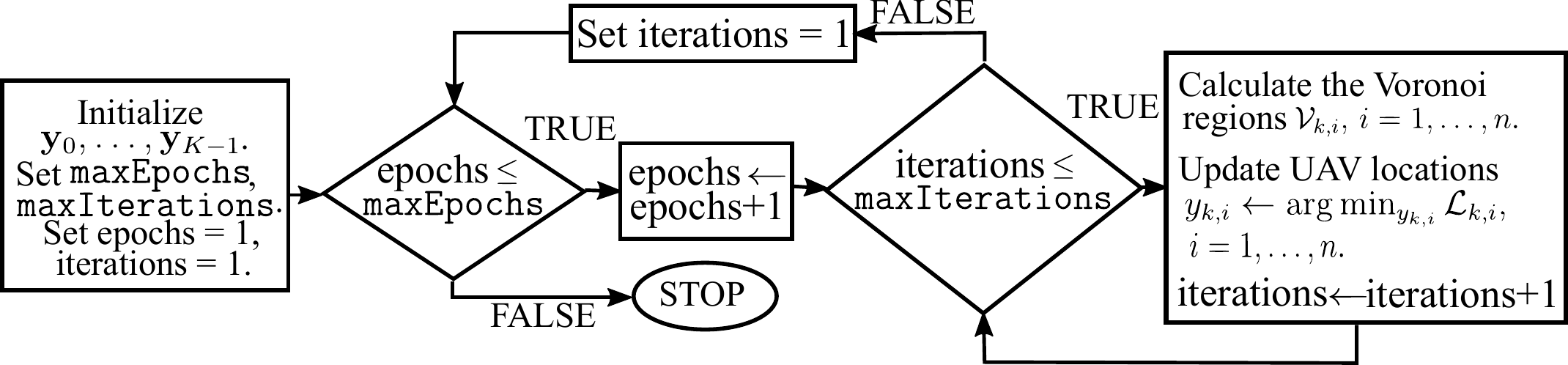}}
\end{center}
\caption{Flowchart of the trajectory optimization algorithm.}
\label{flowchart}
\end{figure*}

\subsection{Minimizing the Lagrangian Cost at a Given Time Instance}
\label{secalgo2}
We now seek a computationally-efficient solution for the $dn$-dimensional optimization problem of minimizing the Lagrangian cost $\mathcal{L}_k$ in (\ref{ellkey}). In other words, we study the optimization problems in Line 3 of Algorithm \ref{algorithm1}. We follow the same decomposition strategy as in Section \ref{secalgo1}. This will lead us to a variant of the Lloyd algorithm of vector quantization \cite{loytoriginal,lbg}. First, for the term $P(\mathbf{y}_{k}, \hat{f}_k)$ in (\ref{ellkey}), we recall from (\ref{avpow}) that 
\begin{align}
\nonumber P(\mathbf{y}_{k}, \hat{f}_k) & = \int_{\mathbb{R}^d} \min_i (\|y_{k,i} - q\|^2 + h^2)^{\frac{r}{2}} \hat{f}_k(q) \mathrm{d}q  \\ & = \sum_{i=1}^n  \int_{\mathcal{V}_{k,i}} (\|y_{k,i} - q\|^2 + h^2)^{\frac{r}{2}} \hat{f}_k(q) \mathrm{d}q,
\end{align}
where $\mathcal{V}_{k,i} \triangleq \{q:\|y_{k,i} - q\| \leq \|y_{k,j} - q\|,\,\forall j\in\{1,\ldots,n\}\}$ is the Voronoi cell of the $i$th UAV at time $k$. Given that the Voronoi cells $\mathcal{V}_{k,i},\,i=1,\ldots,n$ are kept fixed, it follows that for any given UAV index $i\in\{1,\ldots,n\}$, the expression (\ref{ellkey}) depends on $y_{k,i}$ only through the quantity
\begin{multline}
\label{ellkeyay}
 \mathcal{L}_{k,i} \triangleq \frac{1}{K} \int_{\mathcal{V}_{k,i}} (\|y_{k,i} - q\|^2 + h^2)^{\frac{r}{2}} \hat{f}_k(q) \mathrm{d}q + \\  \frac{\ell}{K}   \|y_{k,i} - y_{k-1,i}\| +  \frac{\ell}{K}    \|y_{k,i} - y_{k+1,i}\|.
\end{multline}
The quantity $\mathcal{L}_{k,i}$ can be interpreted as the Lagrangian cost of UAV $i$ at discrete time instance $k$. In general, given that $\mathcal{V}_{k,i},\,i=1,\ldots,n$ are kept fixed, $\mathcal{L}_{k,i}$ is a convex function of $y_{k,i}$ and thus be effectively minimized using any convex optimization method, or gradient descent. For reference, the gradient of $\mathcal{L}_{k,i}$ with respect to $y_{k,i}$ can be calculated to be
\begin{multline}
\frac{\partial \mathcal{L}_{k,i}}{\partial y_{k,i}} = \frac{r}{K} \int_{\mathcal{V}_{k,i}} \frac{(y_{k,i} - q)\hat{f}_k(q)\mathrm{d}q }{(\|y_{k,i} - q\|^2 + h^2)^{1-\frac{r}{2}}} \\
+  \frac{\ell}{K}  \frac{y_{k,i} - y_{k-1,i}}{\| y_{k,i} - y_{k-1,i}\|} + \frac{\ell}{K}  \frac{y_{k,i} - y_{k+1,i}}{\| y_{k,i} - y_{k+1,i}\|}.
\end{multline}
As we shall soon discuss, further simplifications or even closed-form solutions to the problem of minimizing $\mathcal{L}_{k,i}$ are available in certain special cases. Regardless, once each $\mathcal{L}_{k,i},\,i=1,\ldots,n$ are minimized (while keeping $\mathcal{V}_{k,i},\,i=1,\ldots,n$ fixed), the new Voronoi regions $\mathcal{V}_{k,i},\,i=1,\ldots,n$ will be calculated according to the new $y_{k,i},\,i=1,\ldots,n$. Algorithm \ref{algorithm2} summarizes this process of minimizing the cost function $\mathcal{L}_k$ in (\ref{ellkey}). The algorithm proceeds in an iterative manner until a certain maximum number of iterations is reached. By definition, the sequence of costs (as evaluated by (\ref{ellkey})) with Algorithm \ref{algorithm2} is non-increasing, and thus convergent. The convergence proof is identical to that of Algorithm \ref{algorithm1} and is thus omitted for brevity. The flowchart of our algorithm in Fig. \ref{flowchart} combines Algorithms \ref{algorithm1} and \ref{algorithm2} in one unifying diagram.

\begin{algorithm}
\begin{algorithmic}[1]
\vspace{10pt}\STATE Initialize UAV locations $\mathbf{y}_k = [y_{k,1} \cdots y_{k,n}]$. Set $\mathtt{maxIterations}$.
\FOR{$\mathtt{iterations} = 1$ \textbf{to} $\mathtt{maxIterations}$}
\STATE Calculate the Voronoi regions $\mathcal{V}_{k,i},\,i=1,\ldots,n$.
\STATE Update the UAV locations as $y_{k,i} \leftarrow \arg\min_{y_{k,i}} \mathcal{L}_{k,i},\,i=1,\ldots,n$.
\ENDFOR
\end{algorithmic}
\caption{Minimizing $\mathcal{L}_k$}
\label{algorithm2}
\end{algorithm}

\subsection{Minimizing the Lagrangian Cost of a UAV at a Given Time Instance}
\label{secalgo3}
We now consider the minimization of $\mathcal{L}_{k,i}$ for a given UAV index $k$ and time instance $i$. This problem appears in Line 4 of Algorithm \ref{algorithm2}. As we have mentioned in Section \ref{secalgo2}, in general, $\mathcal{L}_{k,i}$ in (\ref{ellkeyay}) can be minimized using gradient descent. Here, we point out that the minimization becomes considerably simpler for the special case $r=2$ of the path loss exponent. In fact, we will also provide a closed-form solution for $r=2$ and one-dimensional networks $d=1$.

Let us first note that for any set $A$ and vector $x$, we have
\begin{align}
& \nonumber \!\!\int_A \|x - q\|^2 f(q)\mathrm{d} q \\ & = \int_A \left(\|x\|^2 - 2x^T q + \|q\|^2\right) f(q)\mathrm{d}q  \\ 
&\nonumber =  \|x\|^2 \left( \int_A\! f(q)\mathrm{d}q\right)  - 2x^T \left( \int_A\! q f(q)\mathrm{d}q\right) + \\ & \qquad\qquad\qquad\qquad\qquad\qquad \qquad\qquad  \int_A \!\|q\|^2 f(q)\mathrm{d}q \\  \nonumber &  = \left( \int_A f(q)\mathrm{d}q\right) \left\|x - \frac{\int_A q f(q)\mathrm{d}q}{\int_A\! f(q)\mathrm{d}q} \right\|^2 + \\ & \label{poqwiepoqiwepoiqweq} \qquad\qquad\qquad\qquad   \int_A \!\|q\|^2 f(q)\mathrm{d}q - \frac{\|\!\int_A q f(q)\mathrm{d}q\|^2}{\int_A f(q)\mathrm{d}q}.
\end{align}
Note that the last two terms do not depend on $x$. When $r=2$, we use the identity in (\ref{poqwiepoqiwepoiqweq}) to rewrite the integral in (\ref{ellkeyay}). Then, by removing the terms that do not depend on $y_{k,i}$, it follows that the minimization of (\ref{ellkeyay}) over $y_{k,i}$ is equivalent to minimizing
\begin{align}
\label{faygenform}
\phi(x) \triangleq \|x - u\| + \|x - v\| + c \|x - w\|^2
\end{align}
over all $x$, where 
\begin{align*}
u & \triangleq y_{k-1,i}, & v & \triangleq y_{k+1,i}, \\ w & \triangleq \frac{\int_{\mathcal{V}_{k,i}} q \hat{f}_k(q) \mathrm{d}q}{\int_{\mathcal{V}_{k,i}} \hat{f}_k(q) \mathrm{d}q},  &  c & \triangleq \frac{1}{\ell} \int_{\mathcal{V}_{k,i}} \hat{f}_k(q) \mathrm{d}q.
\end{align*}

Therefore, when we wish to minimize (\ref{ellkeyay}) (by using gradient descent for example), we can avoid integration over the generally complicated region $\mathcal{V}_{k,i}$ by considering instead the equivalent problem of minimizing (\ref{faygenform}). Let us further note that the domain/search space of minimization of (\ref{faygenform}) is the entire $\mathbb{R}^d$. We can also greatly reduce the size of this search space. For this purpose, we need the following lemma, whose proof can be found in Appendix B. 
\begin{lemma}
\label{trianglelemma}
Let $\mathcal{T} \subset \mathbb{R}^d$ be a triangle with vertices $u,v,w$, including its boundary and interior. For any $x\in\mathbb{R}^d$, there exists $y\in \mathcal{T}$ such that $\|y - a\| \leq \|x - a\|$ for every vertex $a$ of $\mathcal{T}$.
\end{lemma}
We now have the following proposition, which immediately follows from Lemma \ref{trianglelemma}.
\begin{proposition}
A minimizer $x^{\star}$ of (\ref{faygenform}) lies on the triangle $\mathcal{T}$ with vertices $u,v,w$.
\end{proposition}
Therefore, without loss of optimality, we may minimize (\ref{faygenform}) over all $x$ of the form $x = w + \alpha x' + \beta x''$, where $\alpha,\beta \geq 0,\,\alpha+\beta \leq 1$, and $x' = u - w$,  $x'' = v-w$ are triangle edge vectors. Hence, the minimization of  (\ref{faygenform}), which should take place over the entire $\mathbb{R}^d$, can be transformed to a convex optimization over the (two-dimensional) triangle $\alpha,\beta \geq 0,\,\alpha+\beta \leq 1$.

In the special case of one dimension, a minimizer of (\ref{faygenform}) can be found in closed form. This is shown by the following proposition, whose proof can be found in Appendix C. 
\begin{proposition}
\label{exactsolutionprop}
Let $d=1$, and $x^{\star} = \arg\min_{x\in\mathbb{R}^d} \phi(x)$. For a simpler notation, we define the intermediate variables
 $u' \triangleq \min\{u,v\}$, $v'  \triangleq \max\{u,v\}$, $\alpha \triangleq \min\left\{ \left|w - \frac{u+v}{2}\right|, \frac{1}{c}\right\}$. 
 We have
\begin{align}
\label{xstars}
x^{\star} = \left\{\begin{array}{rl} 
w, & w\in[u',v'], \\ 
\max\left\{v', w - \alpha \right\}, & w > v', \\
\min\left\{u', w + \alpha \right\}, & w < u'. \\
\end{array} \right.
\end{align}
\end{proposition}
As a result, the computational complexity of the trajectory optimization algorithm can be greatly reduced for the special case $d=1,r=2$.

\subsection{Implementation and Complexity}
\label{secalgo4}
We envision two possible scenarios in which one can implement our trajectory optimization algorithm in Fig. \ref{flowchart}. In an offline implementation scenario, we may simply run the algorithm on a dedicated server. The final trajectories may then be used by the UAVs on site. In this scenario, the UAVs do not perform any online optimization and simply follow the predetermined paths resulting from the offline optimization on the server. On the other hand, the algorithm also favors online, distributed implementation. For this purpose, suppose that the period of length $T$ is divided into $K$ discrete time slots as before. At discrete time $k$ of a certain epoch, where $k\in\{0,\ldots,K-1\}$, each UAV can calculate its position at time $k$ of the next epoch by communicating with its nearest neighboring UAVs only. In detail, following Line 4 of Algorithm \ref{algorithm2}, at any given discrete time $k$, UAV $i$ first calculates its Voronoi cell $\mathcal{V}_{k,i}$. This can be accomplished by UAV $i$ communicating with its nearest neighboring UAVs only. Each UAV then solves the optimization problem in Line 5 of Algorithm \ref{algorithm2}, possibly by using the closed-form solutions in Section \ref{secalgo3}. This requires only the conditional GT density function on the Voronoi cell $\mathcal{V}_{k,i}$, which can either be made available to UAV $i$ offline, or can be determined via measurements onboard. This online implementation is also distributed in the sense that a given UAV only needs to know the locations of its neighboring UAVs when calculating its location in the next epoch. In particular, a given UAV does not need to track the trajectories of other UAVs.

Let us now discuss the computational complexity of implementing our algorithm with respect to the number of UAVs $n$ in the distributed setting mentioned above. Each UAV needs to execute Lines 3 and 4 of Algorithm \ref{algorithm2} per discrete time slot. Line 3 takes $O(n)$ operations in the worst-case scenario, while Line 4 can be accomplished with $O(1)$ operations. Thus, the total complexity is $O(n)$ operations per UAV per discrete time, or a total of $O(n^2)$  operations per discrete time. Note that, unless the system parameters are among the special cases in Section \ref{secalgo3} (in which case Line 4 can easily be solved in closed form), one has to resort to an iterative algorithm to solve Line 4 of Algorithm \ref{algorithm2}. For a small number of UAVs, such an iterative solution may be the dominating factor in terms of computational complexity.

The advantage of our Lloyd algorithm based approach over the gradient descent based approach of \cite{confversion} is that it favors fast, online, distributed implementation. In fact, comparison of the  numerical simulation results of \cite{confversion} with the ones in the next section reveal that, for all scenarios considered, the performance of our algorithm (in terms of the UAV-GT transmission power tradeoffs) is also no worse than the algorithm in \cite{confversion}. 

\section{Numerical Results} 
\label{secnumerical}

In this section, we provide numerical simulations that verify our analytical results. We first consider the one-dimensional networking scenario in Examples \ref{example1} and \ref{example2}. The corresponding simulation parameters are shown in the first row of Table \ref{table1}. 

\begin{table*}[t]
\caption{Simulation parameters}
\begin{center}
\begin{tabular}{|c|p{.35in}| p{2.2in} |p{.9in}| p{0.6in} |p{.7in}|}
                                         \hline & Region    & Node density & Number of UAVs & Path loss  & UAV altitude \\ \hline
1D network &  [0,3]    & $f_t(q) = (1+3|t|)(q-2+2|t|)^{3|t|}$, $q\in[2-2|t|,3-2|t|],t\in[-1,1]$. & $n\!\in\!\{1,2,4,$ $8,16,32\}$ & $r=2$ & $h=0$  \\ \hline
2D network &  $\mathbb{R}^2$ & $f_t(q) \!=\! \frac{1}{{2\pi\sigma^2}} \exp( \frac{-1}{2\sigma^2}\|q \!-\! [\begin{smallmatrix} 10 \sin 2\pi t  \\ 10 \cos 2\pi t \end{smallmatrix}]\|^2)$, $\sigma = 3+2\sin 2\pi t,\,t\in[0,1]$ & $n\!\in\!\{1,2,4,$ $8,16,32\}$ & $r=3$ & $h=10$ \\ \hline
\end{tabular}
\end{center}
\label{table1}
\end{table*}

\begin{figure*}[h]
\fbox{
    \hspace{-9pt}
    \begin{subfigure}[b]{0.5\textwidth}
        \centering
        \scalebox{0.5}{\includegraphics{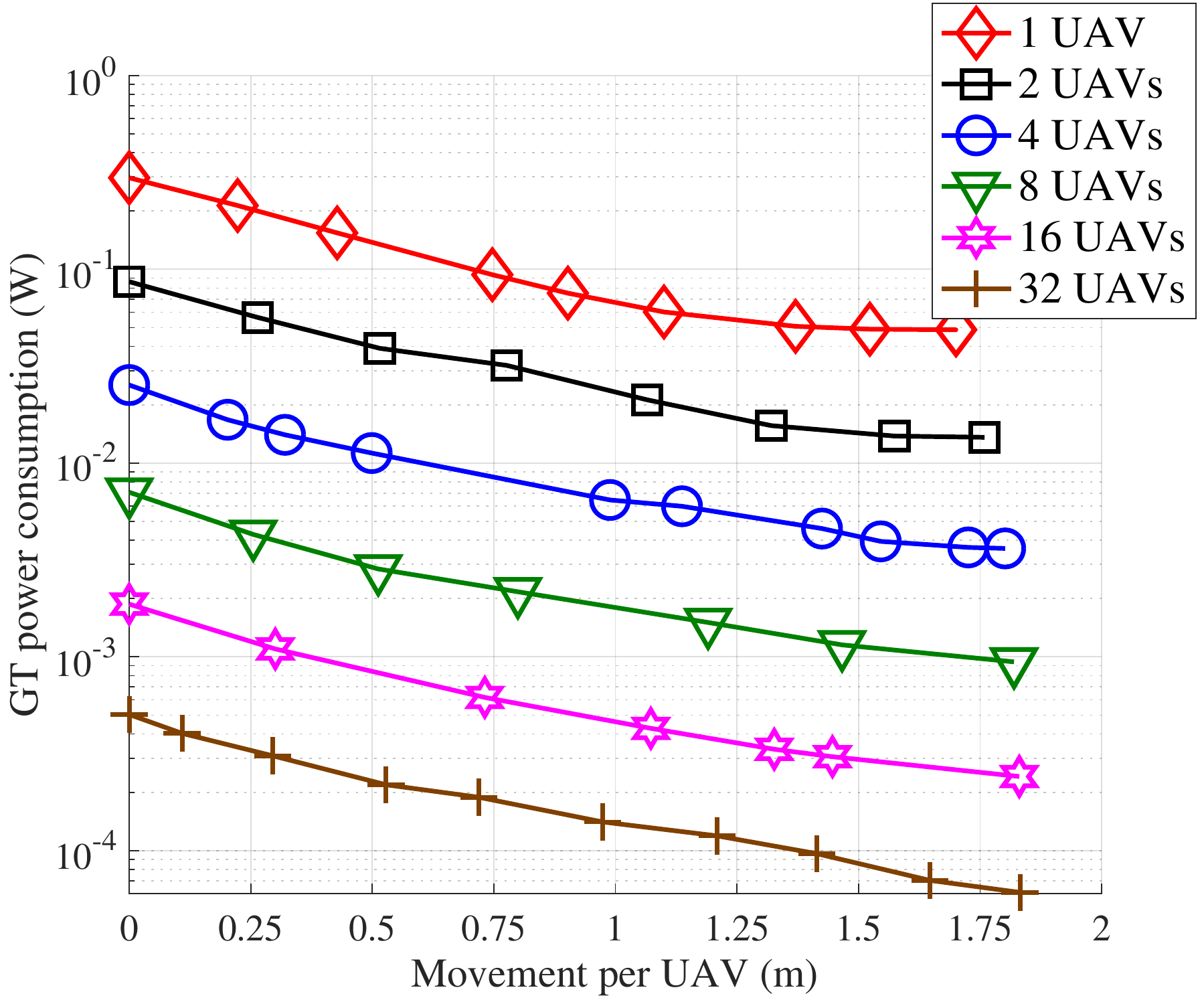}}
        \label{dim1tradeoffsperuav}
    \end{subfigure}%
    \begin{subfigure}[b]{0.5\textwidth}
        \centering
        \scalebox{0.5}{\includegraphics{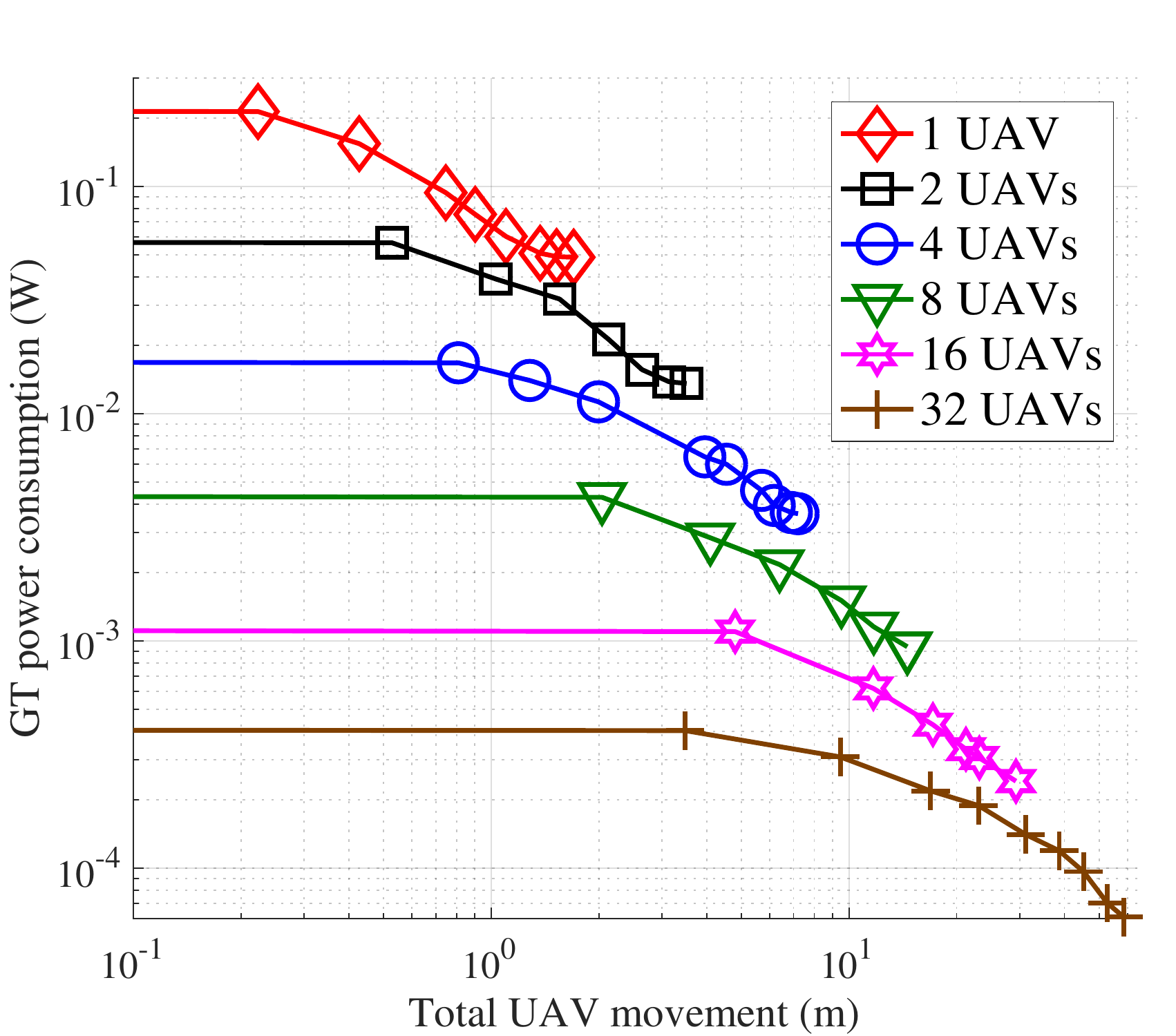}}\vspace{10pt}
         \label{dim1tradeoffstotaluav}
    \end{subfigure}%
    \hspace{-9pt}
    }
    \caption{GT power consumptions for different UAV movements in a one-dimensional network.}
    \label{dim1tradeoffs}
\end{figure*}

In Fig. \ref{dim1tradeoffs}, we show the tradeoff between the UAV movement and the GT power consumption for  different number of UAVs. The two subfigures show the same data points: While in one subfigure, the horizontal axis represents the movement per UAV, in the other subfigure, it represents the total UAV movement. Each marked data point is obtained using the algorithm in Section \ref{secalgogen} for different values of the Lagrange multiplier $\ell$. Throughout the experiments, we have observed that increasing the number of discrete time instances beyond $K=20$ does not significantly improve the continuous-time cost function (\ref{lagranjiyin}). We have thus set a time discretization of $K=20$ for all simulations. Also, we have run the algorithm with different values of the parameter $\mathtt{maxEpochs}$, while $\mathtt{maxIterations}$ is chosen large enough to observe convergence of the overall Lagrangian cost in (\ref{disctimelagrange}). Eventually, we choose the best trajectory that minimizes (\ref{disctimelagrange}). The continuous-time trajectory is reconstructed from the discrete-time trajectory using linear interpolation.

\begin{figure}[h]
\begin{center}
\scalebox{0.5}{\includegraphics{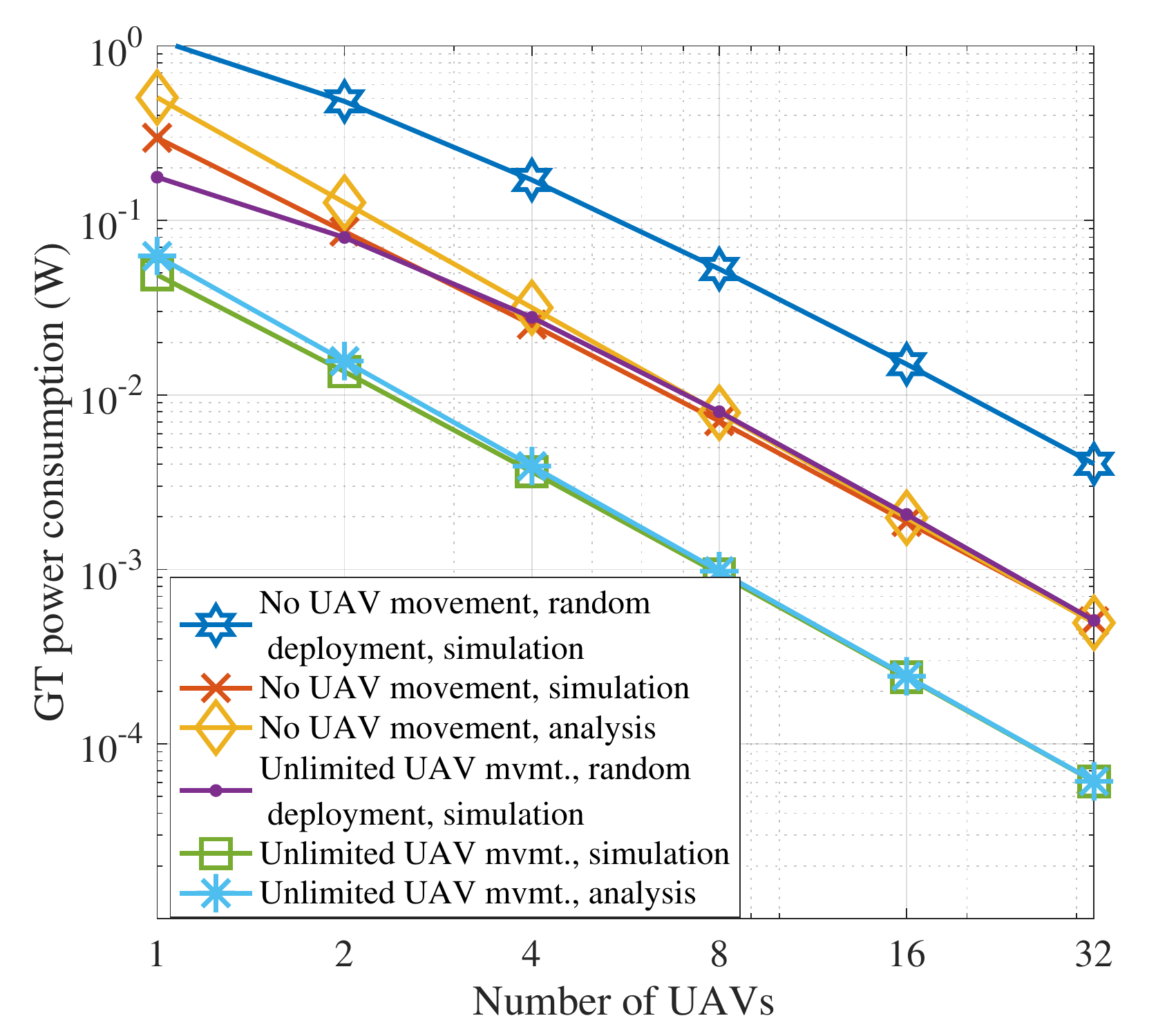}}
\end{center}
\caption{GT power consumptions in extremal cases for a one-dimensional network.}
\label{dim1nouavsvspow}
\end{figure}

We can observe that, doubling the number of UAVs roughly quarters the average GT power consumption for the same amount of distance traveled per UAV. The $O(\frac{1}{n^2})$ decay of the analytical formulae in (\ref{bestpowinonedimzeromvmt}) and (\ref{bestpowinonediminfmvmt}) justify this observation for the special cases of zero and unlimited UAV movement, respectively. The same decay rate can be observed if one instead considers a total movement constraint.

In Fig. \ref{dim1nouavsvspow}, we show the average GT power consumption for different number of UAVs and the extremal scenarios of zero and unlimited UAV movements. The logarithmically-scaled horizontal and the vertical axes represent the number of UAVs and the GT power consumptions, respectively. Note that, the simulation curves for zero and unlimited UAV movements respectively correspond to the vertical coordinates of the leftmost and the rightmost data points in Fig. \ref{dim1tradeoffs}. We can observe that the analytical results in (\ref{bestpowinonedimzeromvmt}) and (\ref{bestpowinonediminfmvmt}) match almost perfectly with the simulations. Due to its asymptotic nature, the analysis is more accurate when the number of UAVs is large. In the same figure, we also show the GT power consumptions given a random UAV deployment algorithm as in \cite{poissonuav}. The algorithm follows the idea of random quantizers \cite{quantization}. Specifically, for the scenario with no UAV movement, we place the UAVs uniformly at random over $[0,3]$, which is the union of the support of the GT densities over the entire period. For the scenario with unlimited UAV movement, we place the UAVs uniformly at random over the support $[2-2|t|,3-2|t|]$ of the GT density at time $t$. The plotted curves are averages over all possible deployments. The two random deployment scenarios are relevant in practice as they demonstrate the achievable performance when only the support of the GT densities are known. As can be observed, for both scenarios of no UAV movement and unlimited UAV movement, the knowledge of the GT density provides roughly an $8$-fold reduction of the GT power consumption.

\begin{figure}[h]
\begin{center}
\scalebox{0.5}{\includegraphics{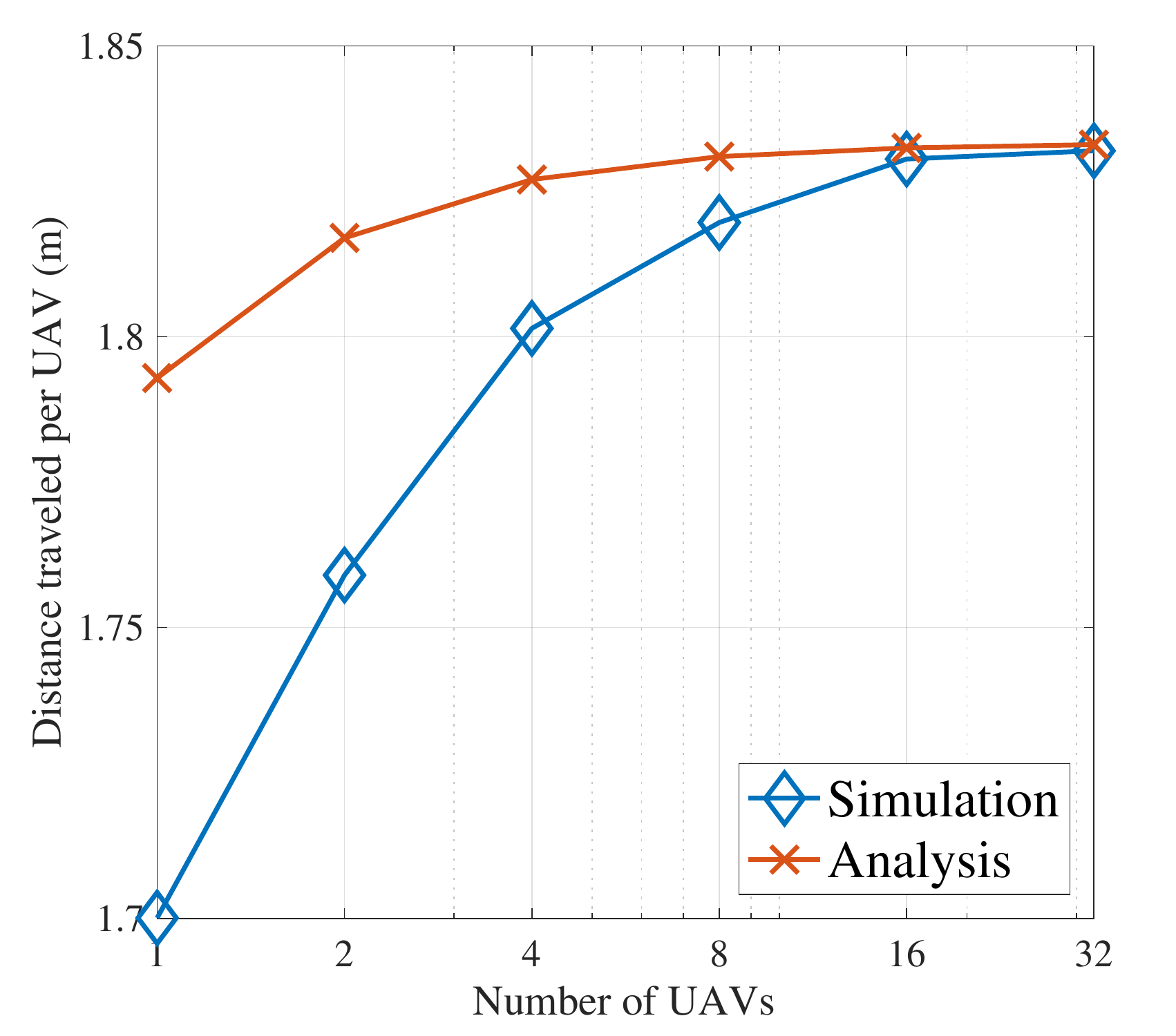}}
\end{center}
\caption{Distance per UAV for different number of UAVs with unlimited movement.}
\label{dim1nouavsvsdist}
\end{figure}

In Fig. \ref{dim1nouavsvsdist}, we show the per-UAV movements for the scenario of (non-random) unlimited UAV movement. The simulation curve corresponds to the horizontal coordinates of rightmost data points in Fig. \ref{dim1tradeoffs}, and the analysis curve corresponds to the formula (\ref{totmvmtformula}). Since the values in the vertical axis of the figure are very close, we can conclude that the analysis matches the simulation very well. In Fig. \ref{dim1nouavsvsdist}, the per-UAV distance grows with the number of UAVs. In this context, one may expect that more UAVs translate to a lower per-UAV movement. In fact, as shown in Fig. \ref{dim1tradeoffs}, given that we consider the same (maximum) GT power consumption, increasing the number of UAVs indeed decreases the per-UAV movement. This is not necessarily the case when the UAVs are instead placed to minimize the GT power consumption without any movement limitations: In this case, given more UAVs, we can afford to place more UAVs to locations with low GT density. If the locations of such low density regions are rapidly varying over time, the end result is a larger per-UAV movement.

\begin{figure}[h]
\begin{center}
\scalebox{0.5}{\includegraphics{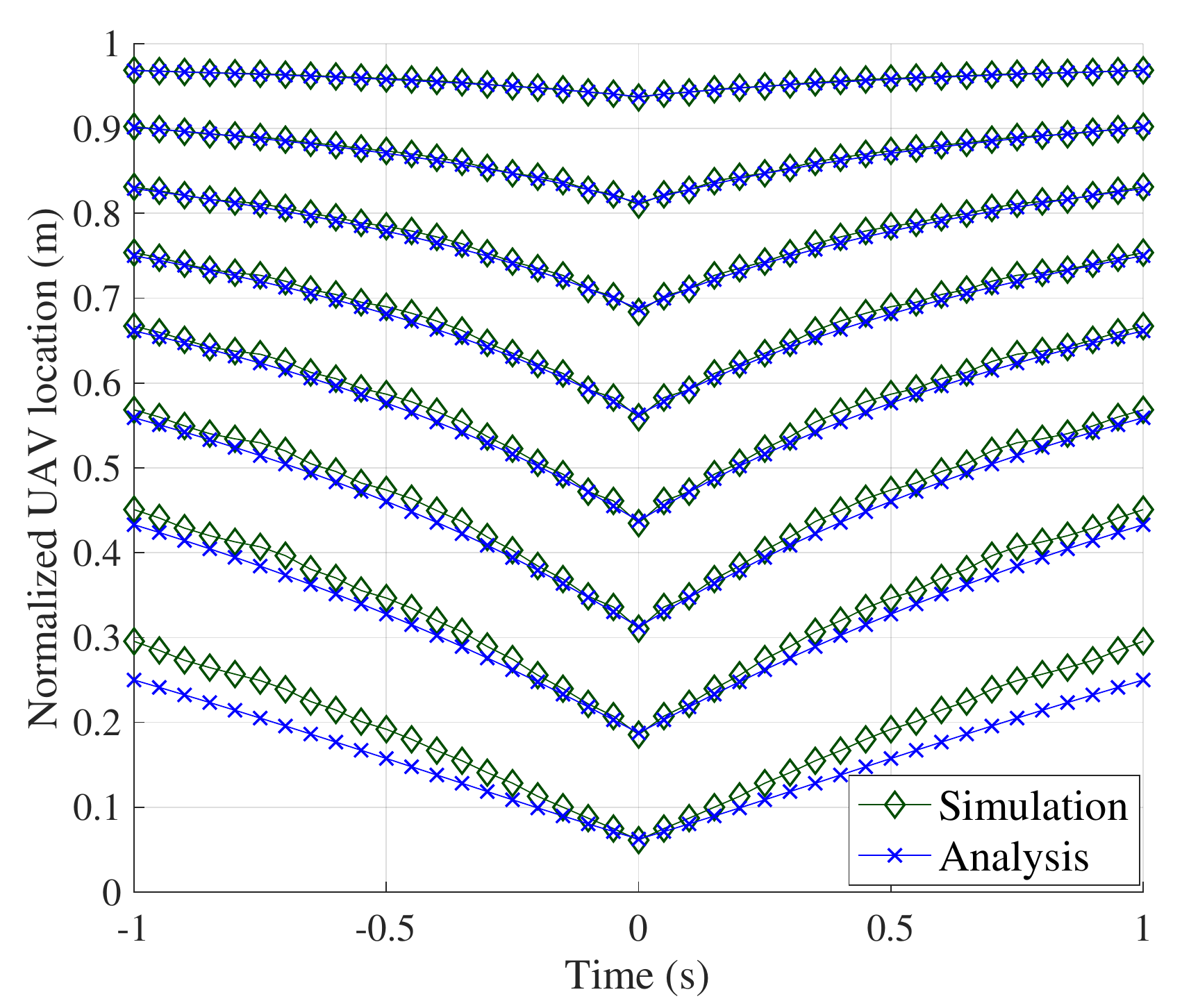}}
\end{center}
\caption{Trajectories of 8 UAVs in a one-dimensional network.}
\label{dim1trajfig}
\end{figure}

Example \ref{example2} also provides the optimal trajectories (\ref{optuavtrajsjsjss}) for the unlimited movement scenario. In Fig. \ref{dim1trajfig}, we compare these analytical trajectories with the trajectories that are obtained numerically for the special case of $8$ UAVs. The horizontal and the vertical axes represent the time, and the UAV locations, respectively. Each curve represents the trajectory of one UAV. We have normalized both the analytical and the simulation trajectories by subtracting the time-varying drift $2-2|t|$ of the density function. We can observe that, for any UAV index, the analysis matches the simulation very well.

\begin{figure}[h]\begin{center}
\scalebox{0.5}{\includegraphics{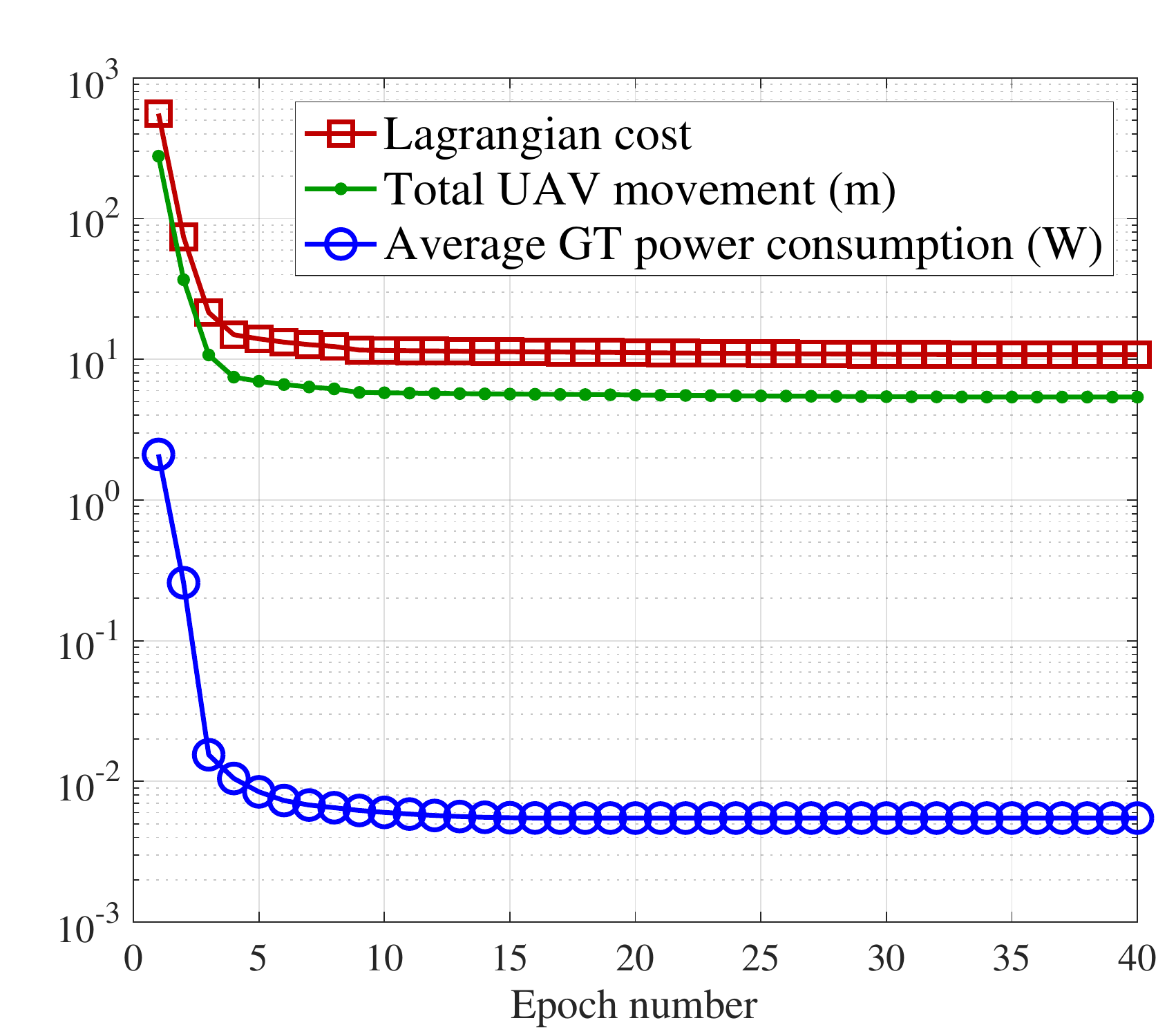}}
\end{center}
\caption{Convergence for an example run of the trajectory optimization algorithm.}
\label{convergencefig}
\end{figure}

In Fig. \ref{convergencefig}, we show the convergence of the Lagrangian cost in (\ref{disctimelagrange}) for an example run of our trajectory optimization algorithm. In the example run, we have considered a Lagrange multiplier of $\ell = 2$, which provided a GT power consumption of around $5.5\times 10^{-3}$ W for a total UAV movement of around $5.4$ m. We can observe that the Lagrangian cost decays very rapidly in the first few epochs, and converges to a value of approximately $10.8$.

As an example of two-dimensional ($d=2$) dynamic deployment, we consider the parameters in the second row of Table \ref{table1}. By Theorems \ref{t1} and \ref{t2}, the asymptotic GT power consumptions are 
\begin{align}
\label{qpipoiqpwoeipqowe1}
Q^{\star}(0) & =  1000+\frac{25}{6\sqrt{3}}\frac{\|\overline{f}\|_{\frac{1}{2}}}{n} + o\left(\frac{1}{n}\right), \mbox{ and }\\
\label{qpipoiqpwoeipqowe2}
Q^{\star}(\infty) & =  1000+\frac{25}{6\sqrt{3}}\frac{\int_0^1\|f_t\|_{\frac{1}{2}} \mathrm{d}t }{n} + o\left(\frac{1}{n}\right),
\end{align}
for the cases of zero and unlimited UAV movement, respectively. For (\ref{qpipoiqpwoeipqowe1}), we can obtain $\|\overline{f}\|_{\frac{1}{2}} \approx 908.16$ via numerical integration. For the case of unlimited movement in (\ref{qpipoiqpwoeipqowe2}), after some calculus, we can obtain $\int_0^1\|f_t\|_{\frac{1}{2}} \mathrm{d}t = 88\pi$ in closed form. 

As the asymptotic expressions (\ref{qpipoiqpwoeipqowe1}) and (\ref{qpipoiqpwoeipqowe2}) also show, the choice of parameters $r=3$ and $h=10$ imply that the GT power consumption is at least $h^r = 1000$ regardless of the constraints on the total UAV movement. For a clear illustration of results, we thus normalize the GT transmission power by subtracting 1000 from the true GT transmission power. 

\begin{figure*}[h]
\fbox{
    \hspace{-5pt}
    \begin{subfigure}[b]{0.5\textwidth}
        \centering
        \scalebox{0.5}{\includegraphics{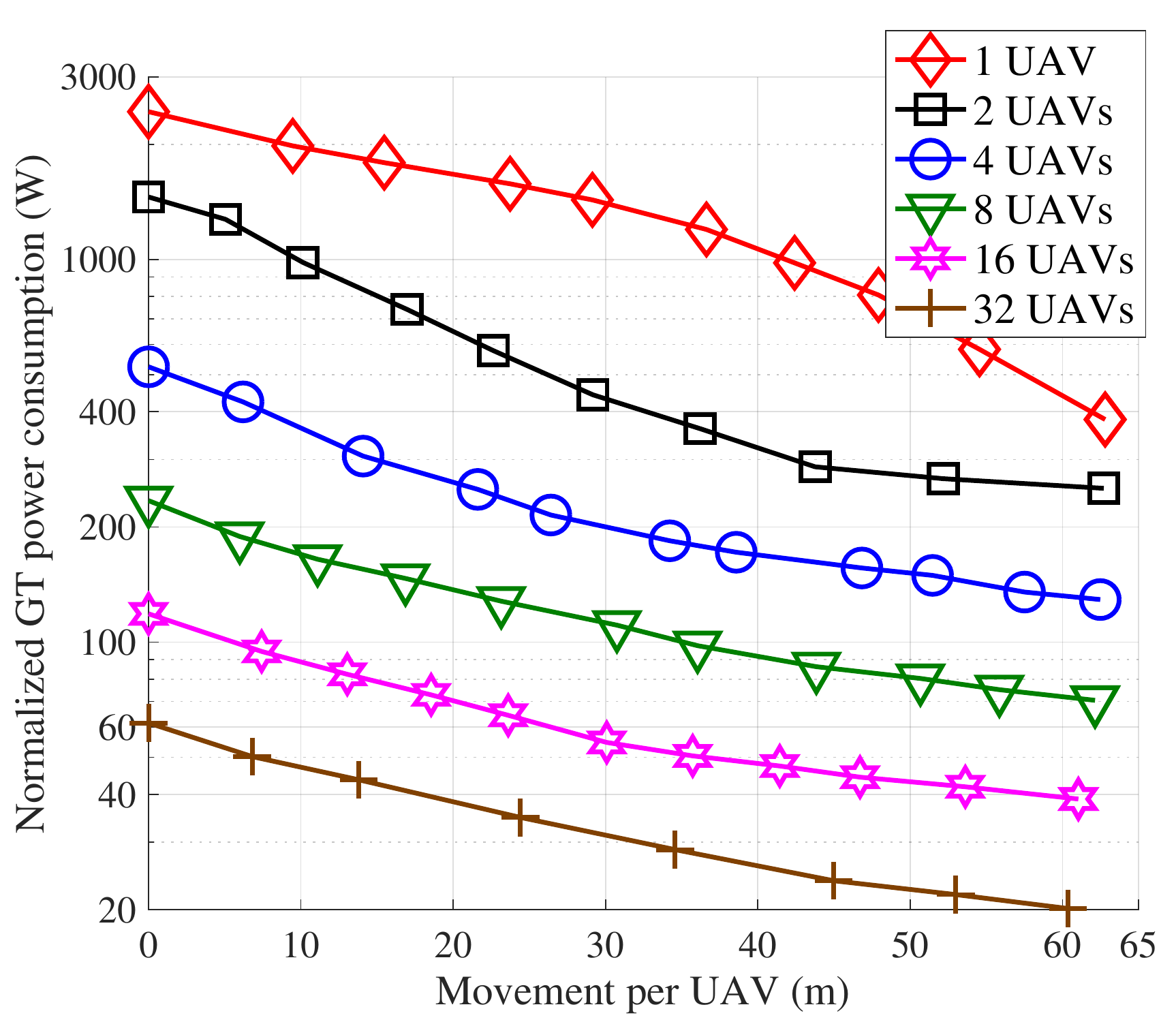}}
        \label{dim2tradeoffsperuav}
    \end{subfigure}%
    \begin{subfigure}[b]{0.5\textwidth}
        \centering
        \scalebox{0.5}{\includegraphics{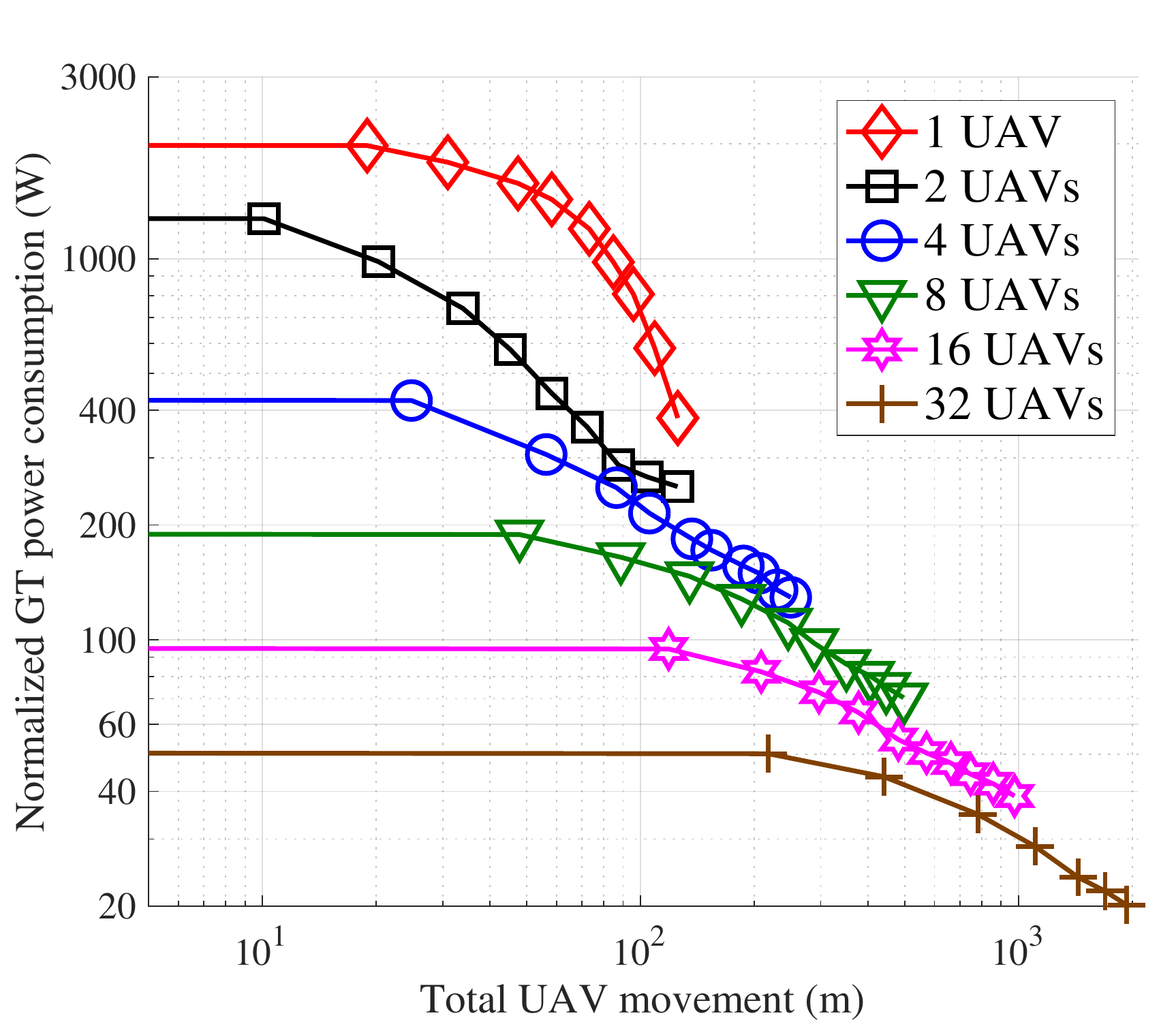}}
         \label{dim2tradeoffstotaluav}
    \end{subfigure}%
      \hspace{-15pt}}
    \caption{GT power consumptions for different UAV movements in a two-dimensional network.}
    \label{dim2tradeoffs}
\end{figure*}

In Fig. \ref{dim2tradeoffs}, we show the tradeoff between the  per-UAV movement and the normalized GT power consumption for different number of UAVs. Unlike the case of the one-dimensional network shown earlier, for a fixed per-UAV movement, doubling the number of UAVs roughly only halves (instead of quartering) the normalized GT power consumption. The asymptotic expression in (\ref{qpipoiqpwoeipqowe1}) formally verifies this observation for the special case of zero movement.

\begin{figure}[h]
\begin{center}
\scalebox{0.5}{\includegraphics{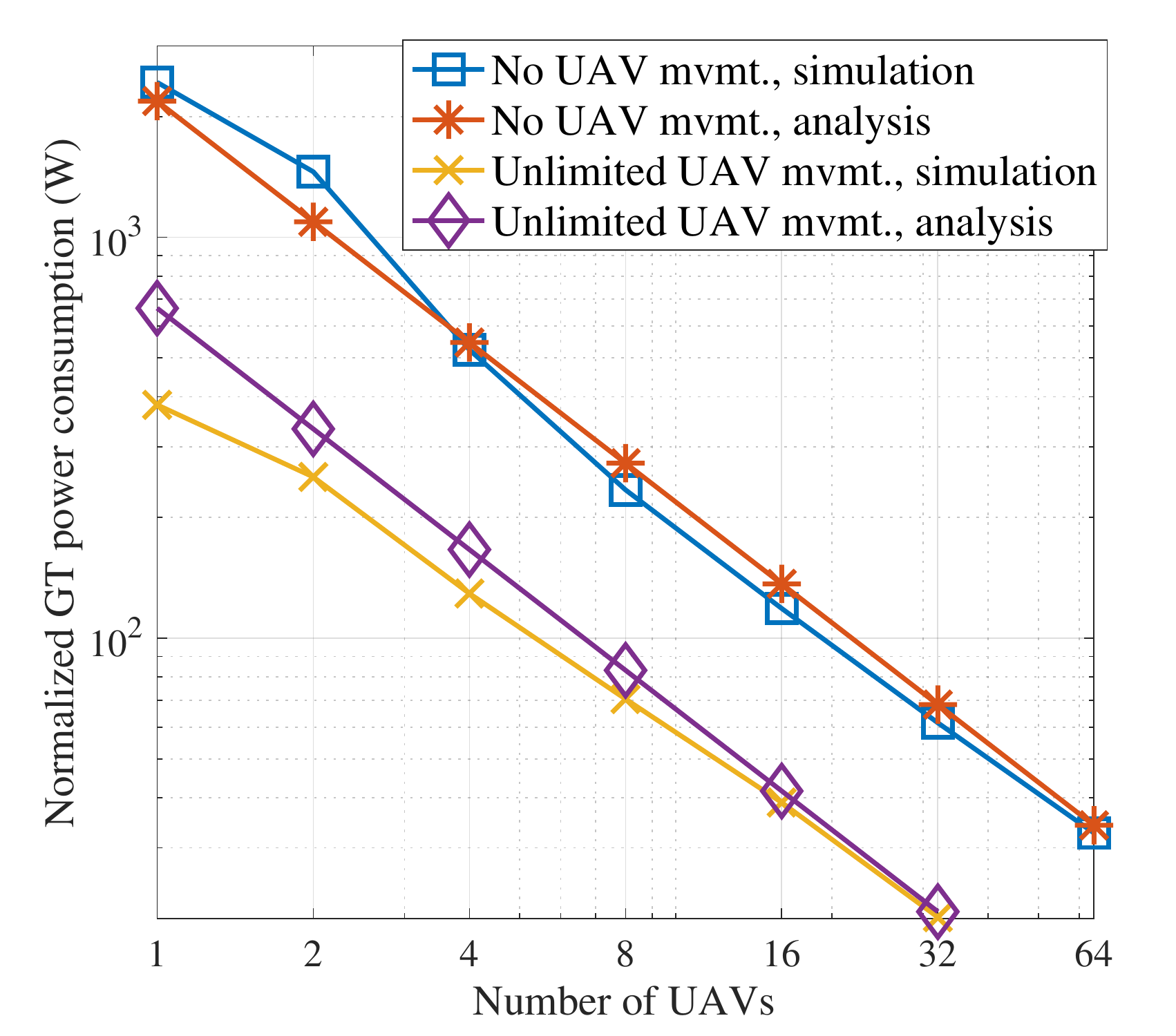}}
\end{center}
\caption{GT power consumptions in extremal cases for a two-dimensional network.}
\label{dim2nouavsvspow}
\end{figure}

In Fig. \ref{dim2nouavsvspow}, we show the average GT power consumption for different number of UAVs in the zero and unlimited UAV movement scenarios. We can observe that the analytical results in (\ref{qpipoiqpwoeipqowe1}) and (\ref{qpipoiqpwoeipqowe2}) match very well with the simulations. In particular, for the case of no UAV movement, the mismatch between the analysis and the simulation increases after $4$ UAVs, and decreases after $16$ UAVs. The reason for the mismatch is that our analysis is only asymptotically tight for a large number of UAVs. In this context, even though the analysis will provide an asymptotically tight approximation on the simulation results, the amount of mismatch for a moderate number of UAVs is also more pronounced as compared to the case of the one-dimensional network. The reason is the smaller amount of UAVs per dimension. One encounters the same phenomenon in the performance analysis of general vector quantizers.

\begin{figure*}[h]
\fbox{
    \hspace{-9pt}
    \begin{subfigure}[b]{0.33\textwidth}
        \centering
        \scalebox{0.35}{\includegraphics{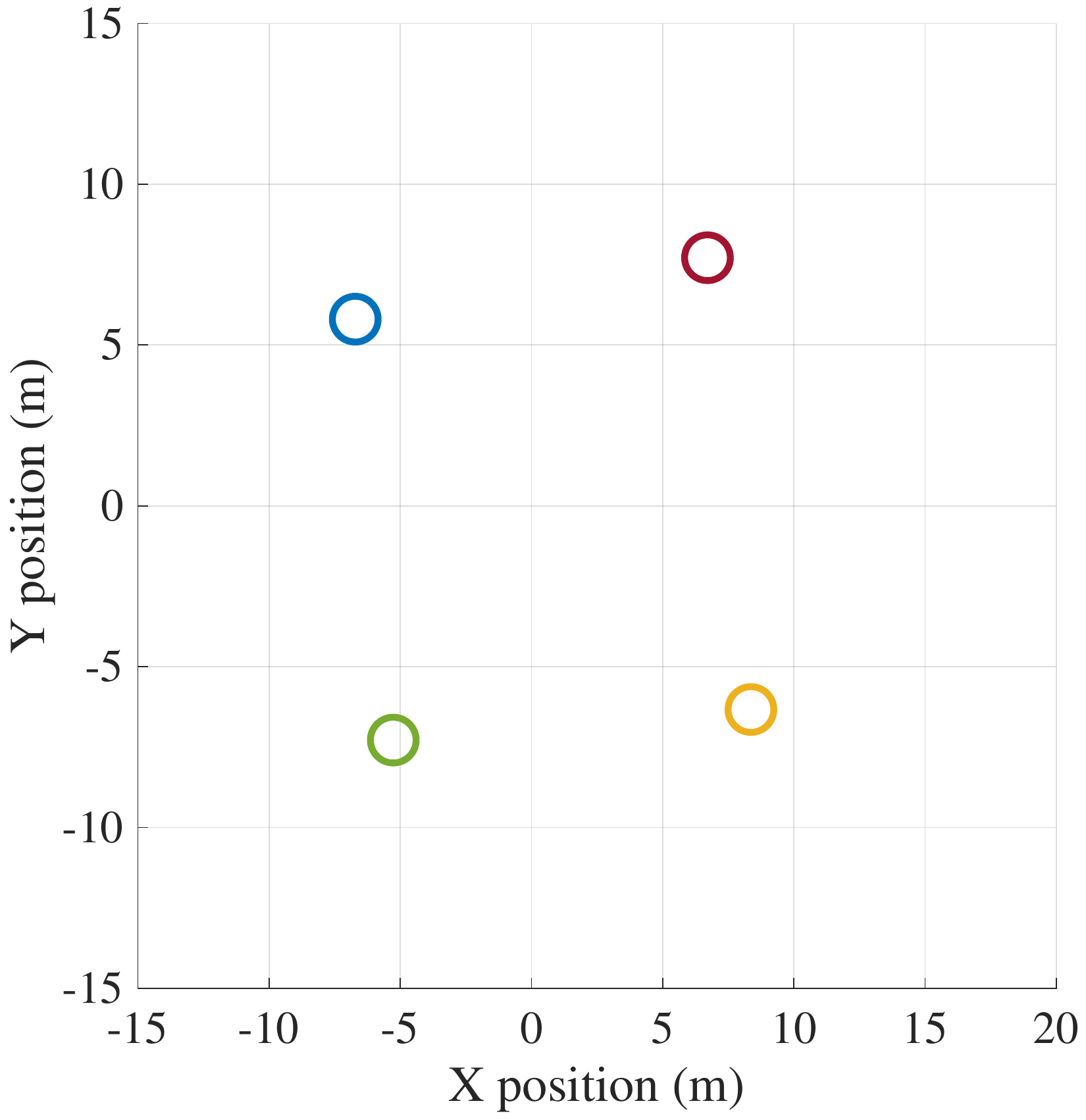}}
        \caption{No UAV movement}
    \end{subfigure}%
        \begin{subfigure}[b]{0.33\textwidth}
        \centering
        \scalebox{0.35}{\includegraphics{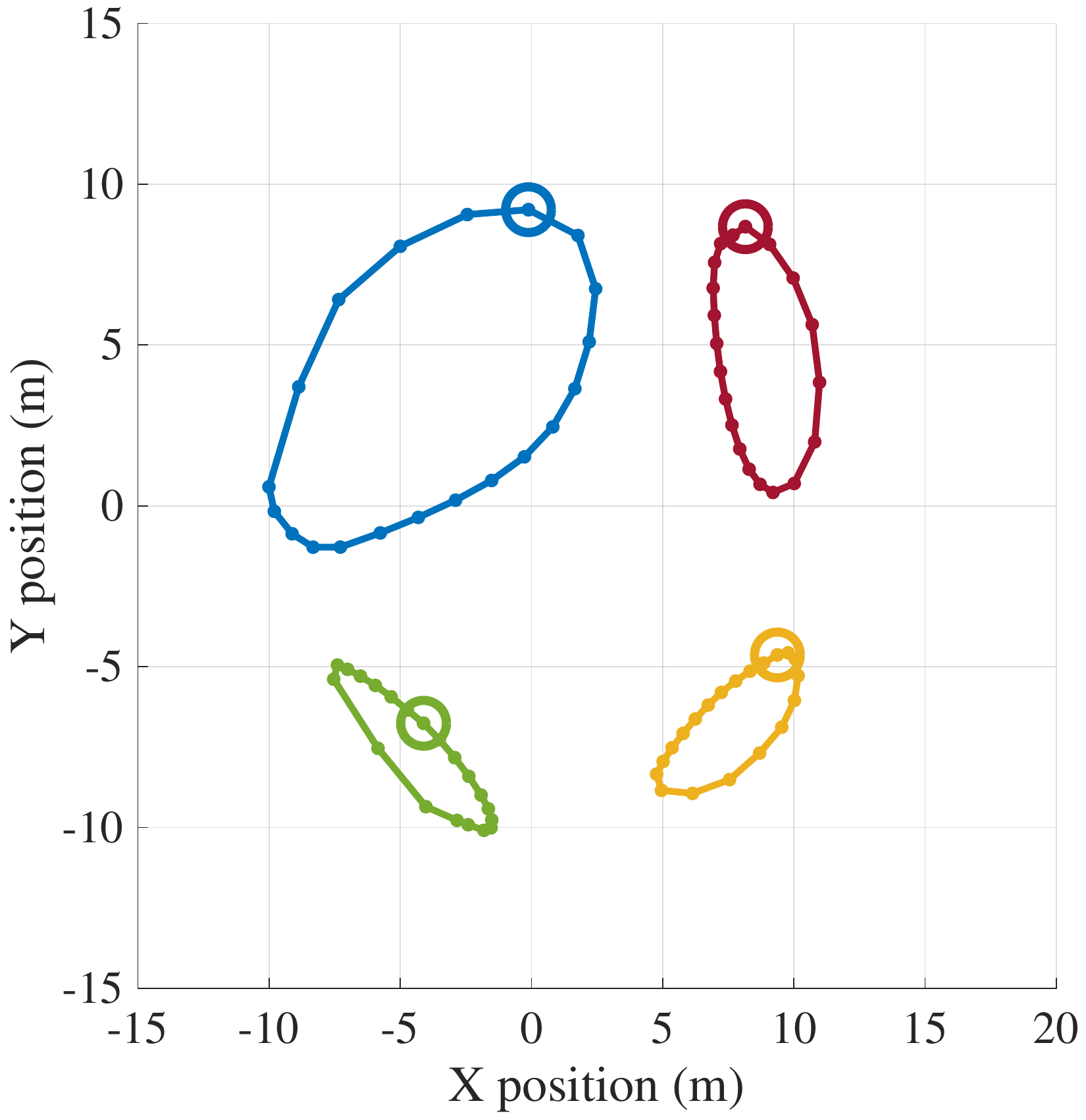}}
        \caption{Moderate UAV movement}
    \end{subfigure}%
        \begin{subfigure}[b]{0.33\textwidth}
        \centering
       \scalebox{0.35}{ \includegraphics{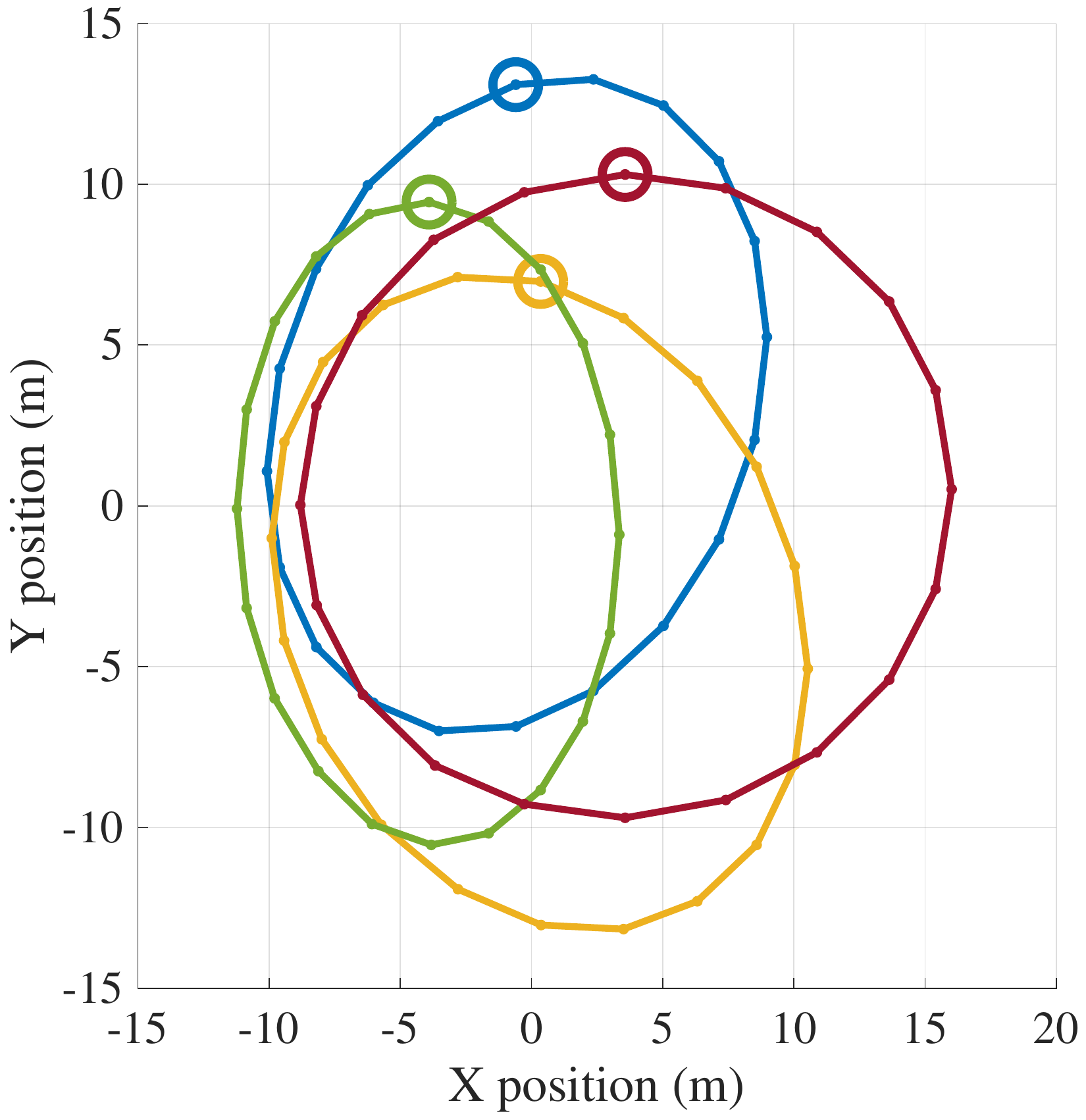}}
       \caption{Unlimited UAV movement}
    \end{subfigure}
    \hspace{-5pt}}
    \caption{Sample trajectories with 4 UAVs. The dimensionalities of the axes are meters.}
    \label{dim2trajfig}
\end{figure*}

Finally, in Fig. \ref{dim2trajfig}, we show the optimized UAV trajectories for the three different scenarios of zero, moderate, and unlimited UAV movements. The moderate movement scenario is designed with a Lagrange multiplier of $\ell = \frac{3}{2}$, and achieves the data point with a per-UAV traveled distance of approximately $21$ in Fig. \ref{dim2nouavsvspow}. The $t=0$ positions of each trajectory are marked with a circle.  Each point on the trajectory corresponds to one discrete time instance that is optimized via the trajectory optimization algorithm. All UAVs travel ``clockwise.'' 

\section{Extensions}
\label{secextensions}
In the previous sections, we have studied the UAV deployment and trajectory optimization problem for a fixed-rate variable-power system. Also, we have considered  a simple line of sight channel model without fading and ignored the effects of multi-user interference. In this section, we consider the extensions of our results to different scenarios.
\subsection{Probabilistic Line of Sight Channel Model}
\label{uavlossection}
We first extend our results to the probabilistic line of sight channel model \cite{uavlospaper, uavlospaper2}. To introduce the model, suppose that a GT at $q$ wishes to communicate with its closest UAV, which is within distance $\min_i \|q - x_i\|$. Then, the transmitted signal of the GT undergoes the same line of sight model of Section \ref{secstaticdeployment} with probability $p_{\mathrm{LOS}} \triangleq \frac{1}{1+c \exp(-b(\theta - c))}$, where $b,c>0$ are constants and $\theta = \tan^{-1} \frac{h}{\min_i \|q - x_i\|}$. On the other hand, with probability $1 - p_{\mathrm{LOS}}$, the GT signal undergoes an extra attenuation of $\delta < 1$. It follows that, given UAV deployment $\mathbf{x}$, and GT density $f$, the average GT power consumption for reliable communication is given by
\begin{multline}
P'(\mathbf{x},f) \triangleq \int_{\mathbb{R}^d} \left(\min_i \|x_i - q\|^2 + h^2\right )^{\frac{r}{2}} \\ \left( p_{\mathrm{LOS}} + \frac{1}{\delta}(1-p_{\mathrm{LOS}})\right)  f(q)\mathrm{d}q.
\end{multline}
This expression is the analogue of (\ref{avpow}) for the probabilistic line of sight channel model. For a large number of UAVs, by a Taylor series expansion, we obtain
\begin{multline}
\label{qowepqowiepqwe}
P'(\mathbf{x},f) = h^r(d+ \tfrac{1-d}{\delta}) + \\ h^r(\tfrac{1}{\delta}  -  1) bcd^2 \int_{\mathbb{R}^d} \min_i \|x_i - q\| f(q)\mathrm{d}q
  + \\ \int_{\mathbb{R}^d} o\left(\min_i \|x_i - q\| \right) f(q)\mathrm{d}q,
\end{multline}
where $d = \frac{1}{1+c \exp(-b(\frac{\pi}{2}-c))+c}$. The only differences between (\ref{qowepqowiepqwe}) and (\ref{avpowwewew}) are in the constants. All of our asymptotic results thus easily extend to the probabilistic line of sight model.
\subsection{Variable-Rate Fixed-Power Systems}
\label{secvarratefixpow}
Let us now discuss variable-rate fixed-power systems. In this case, each GT transmits with a fixed power $P$, resulting in the achievable average rate (in nats/sec/Hz)
\begin{align}
\label{oqjwpoejqpowje}
R(\mathbf{x},f) \!\triangleq\! \int_{\mathbb{R}^d}\! \log\!\left( 1\!+\! \frac{P}{(\min_i \|x_i \!-\! q\|^2 \!+\! h^2 )^{\frac{r}{2}}} \right)\! f(q)\mathrm{d}q,
\end{align}
as the variable-rate analogue of (\ref{avpow}). For a large number of UAVs, a Taylor expansion yields
\begin{multline}
\label{oqjwpoejqpowje2}
R(\mathbf{x},f)  =   \log_2\!\left(1+ \frac{P}{h^r} \right) - \\ \frac{rP/\log 2}{2 h^2(P+h^r)}\int_{\mathbb{R}^d} \min_i \|x_i - q\|^2 f(q)\mathrm{d}q
  + \\ \int_{\mathbb{R}^d}o\left(\min_i \|x_i - q\|^2\right) f(q)\mathrm{d}q.
\end{multline}
Similarly, comparing with (\ref{avpowwewew}), the only differences are in the constants. Thus, our results also extend to variable-rate systems in a straightforward manner.
\subsection{Effects of Fading and Interference in Uplink or Downlink Communications}
\label{interferencesec}
We now consider a UAV-based network that takes into account the effects of fading and interference for either uplink or downlink communications. Specifically, we study a scenario where the $n$ UAVs form a distributed base station with $N$ antennas, and each UAV has $\frac{N}{n}$ antennas. Consider the case of uplink communications where $m$ single-antenna users simultaneously wish to communicate with the UAVs; the downlink case results in the same cost functions and its analysis is thus identical. In particular, given $j\in\{1,\ldots,m\}$, User $j$ wishes to communicate the complex Gaussian symbol $s_j \sim \mathcal{CN}(0,1)$ to the UAVs by transmitting the signal $s_j \sqrt{P}$ over its single antenna. Given $i\in\{1,\ldots,n\}$, $j\in\{1,\ldots,m\}$, and $l\in\{1,\ldots,\frac{N}{n}\}$, let $h_{jil}\in\mathbb{C}$ be the channel gain between User $j$ and Antenna $l$ of UAV $i$. We assume that $h_{jil} \sim \mathcal{CN}\left(0,(h^2 + \|x_i - q_j\|^2)^{-\frac{r}{2}}\right)$, 
where $q_j\in\mathbb{R}^d$ is the location of User $j$. 

The channel input-output relationships are 
$y_{il} = \sum_{j=1}^m h_{jil} s_j \sqrt{P} + \tau_{i l}$, where $\tau_{il} \sim \mathcal{CN}(0,1)$ is the noise at the $l$th antenna of UAV $i$. All the channel gains, noises, and the data symbols are assumed to be independent. Here, we consider the massive MIMO regime where the total number of UAV antennas $N$ grows to infinity. In such a scenario, the achievable rate for a generic user at location $q$ is given by $\log_2( 1+  \sum_{i = 1}^{n} \frac{P}{(h^2 + \|x_i - q\|^2)^{r/2}} )$ bits/sec/Hz \cite{koyuncu0}. Given user density $f$ and UAV deployment $\mathbf{x}$, the average achievable rate of a user is then
\begin{align}
\label{odqwodiqwdqwdqw}
\!\!\widetilde{R}(\mathbf{x},f) \!\triangleq\!  \int_{\mathbb{R}^d} \!\log_2 \left( 1\!+\!   \sum_{i = 1}^{n} \frac{P}{(h^2 \!+\! \|x_i \!-\! q\|^2)^{\frac{r}{2}}} \right)\! f(q)\mathrm{d}q.
\end{align}
Consider now the static or the dynamic deployment problem for the cost function in (\ref{odqwodiqwdqwdqw}). For large path loss exponents, Lemma 1 of \cite{koyuncu0} can be used to obtain the approximation
\begin{align}
\label{oqiwoeqiweqw}
\!\!\widetilde{R}(\mathbf{x},f)\! \simeq \! \int_{\mathbb{R}^d} \log_2\! \left( 1\!+\!   \frac{P}{(h^2 \!+\! \min_i \|x_i\! -\! q\|^2)^{\frac{r}{2}}} \right) f(q)\mathrm{d}q\!\!
\end{align}
that holds for any non-degenerate deployment with $x_i \neq x_j \iff i \neq j$. Noting that (\ref{oqiwoeqiweqw}) and (\ref{oqjwpoejqpowje}) are equal, as discussed after (\ref{oqjwpoejqpowje}), we can use the Taylor series expansion in (\ref{oqjwpoejqpowje2}) to reduce the problem to the one studied in Section \ref{secdegenerate}. All of our asymptotic results then extend to a scenario with fading and interference in a straightforward manner.

\section{Conclusions} 
\label{secconclusions}
We have studied the optimal deployment and relocation of UAV networks. For static networks without any GT density variations, we have found the asymptotically optimal UAV locations that minimize the average GT power consumption or maximize the GT data rate. We have also provided analytical and numerical methods for dynamic UAV deployment where the GT density varies over time. In particular, we have found the asymptotically optimal UAV trajectories for one-dimensional networks and an unlimited UAV movement. We have also introduced a trajectory optimization algorithm for finding good trajectories for moderate UAV movement constraints. 

\section*{Acknowledgements}
The authors would like to thank Raheleh Khodabakhsh and Nitin Surya for their invaluable feedback on an earlier version \cite{confversion} of this paper.

\medskip
\begin{center}
\textsc{Appendix A: Proof of Proposition \ref{onedimuniprop}}
\end{center}

First, note that if $g$ is monotonically non-decreasing, $A\subset\mathbb{R}$, and $x\in\mathbb{R}$, we have 
$\int_A g(\|x - q\|)\mathrm{d}q \geq \int_{B} g(\|q\|)\mathrm{d}q$, where $B$ is the origin-centered interval with the same measure as $A$.  In particular, for $g(u) = (u^2 + h^2)^{\frac{r}{2}}$, we obtain   $\int_A (\|x - q\|^2 + h^2)^{\frac{r}{2}} \mathrm{d}q \geq  h(\mu(A))$, 
where $h(\nu) \triangleq \int_{-\frac{1}{2} \nu}^{\frac{1}{2} \nu} (u^2 + h^2)^{\frac{r}{2}} \mathrm{d}u = 2\int_{0}^{\frac{1}{2} \nu} (u^2 + h^2)^{\frac{r}{2}} \mathrm{d}u$, and $\mu(A)$ is the Lebesgue measure of $A$. By differentiation, we can show that $h(\nu)$ is concave in $\nu$. Now, let $V_i \triangleq \{q:\|q-x_i\| \leq \|q - x_j\|,\,\forall j\},\,i=1,\ldots,n$ denote the Voronoi cells that are generated by $x_1,\ldots,x_n$. We have
\begin{multline}
 P(\mathbf{x}, f)   = \sum_{i=1}^n \int_{V_i} (\|x_i - q\|^2 + h^2)^{\frac{r}{2}}\mathrm{d}q 
 \geq \sum_{i=1}^n h(\mu(V_i))  \\ \!\!\geq n h\left( \frac{1}{n} \sum_{i=1}^n \mu(V_i) \right) 
   \!=\! n h\!\left( \!\frac{1}{n} \!\right)  \! = \!2n \int_{0}^{\frac{1}{2n}} \!\!(u^2 \!+\! h^2)^{\frac{r}{2}} \mathrm{d}u.\!\!
\end{multline}
The  second inequality follows from the concavity of $h(\cdot)$. It can easily be verified that the last expression equals $P(\mathbf{x}_u,f)$. This concludes the proof.

\begin{center}
\textsc{Appendix B: Proof of Lemma \ref{trianglelemma}}\vspace{-5pt}
\end{center}

\begin{figure}[h]
\begin{center}
    \begin{subfigure}[b]{0.24\textwidth}
        \centering
        \scalebox{0.3}{\includegraphics{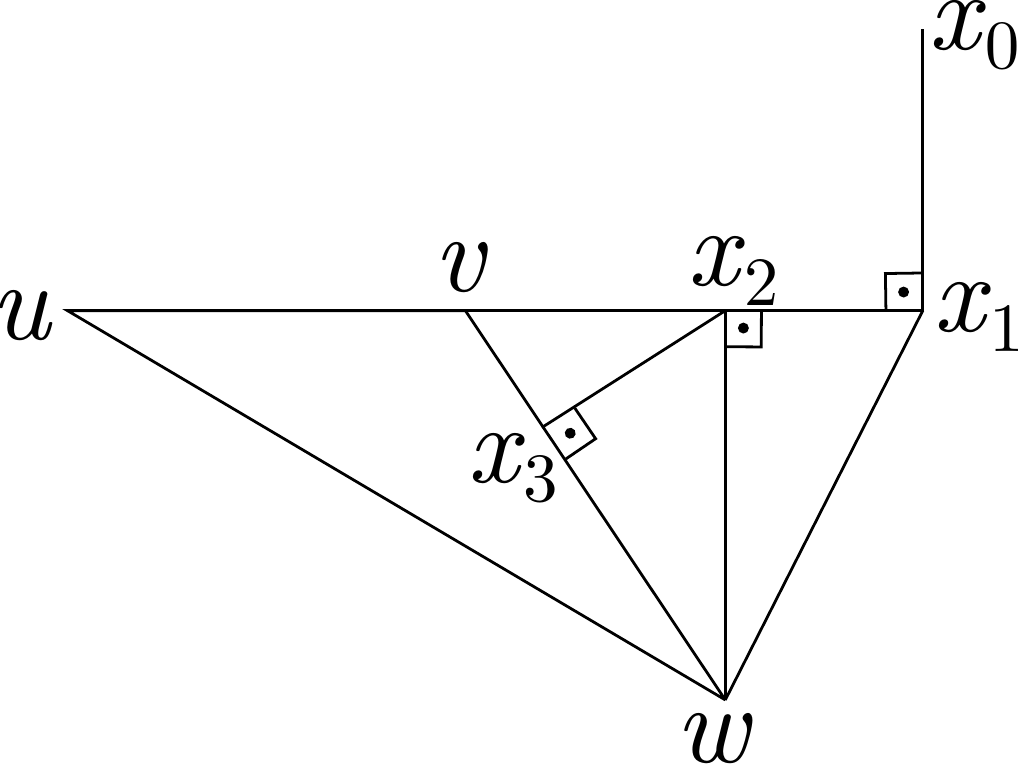}}
        \caption{The first case}
        \label{trproofcase2}
    \end{subfigure}
        \begin{subfigure}[b]{0.24\textwidth}
        \centering
        \scalebox{0.3}{\includegraphics{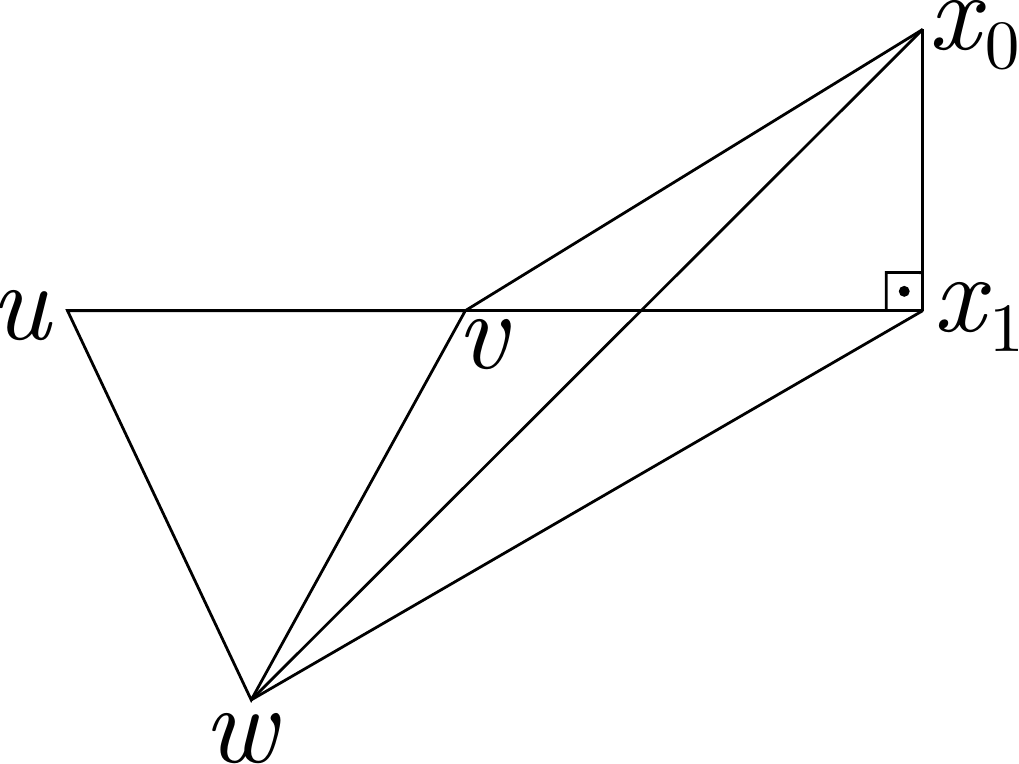}}
        \caption{The second case}
        \label{trproofcase1}
    \end{subfigure}%
    \end{center}
    \caption{Two cases for the proof of Lemma \ref{trianglelemma}.}
\end{figure}

Let  ``$y \ll x$'' denote the conditions $\|y - u\| \leq \|x - u\|$, $\|y - v\| \leq \|x - v\|$, and $\|y - w\| \leq \|x - w\|$.
If $x\in \mathcal{T}$, we set $y = x$, and the proof is complete. Otherwise,  let $x_0$ be the projection of $x$ on the two-dimensional subspace that contains $\mathcal{T}$. We have $x_0 \ll x$ by the Pythagorean inequality. If $x_0 \in \mathcal{T}$, the lemma then follows with $y = x_0$. Otherwise, by appropriate translations of $u,v,w,x_0$, we may assume $u = \left[\begin{smallmatrix} 0 \\ 0 \end{smallmatrix}\right]$, $v = \left[\begin{smallmatrix} v_1 \\ 0 \end{smallmatrix}\right]$, $w = \left[\begin{smallmatrix} w_1 \\ w_2 \end{smallmatrix}\right]$, and $x_0 =  \left[\begin{smallmatrix} x_{01} \\ x_{02} \end{smallmatrix}\right]$, where $v_1,x_{01},x_{02} \geq 0$, $w_2 \leq 0$, and $w_1 \in \mathbb{R}$. Now, let $x_1 =  \left[\begin{smallmatrix} x_{01} \\ 0 \end{smallmatrix}\right]$. The geometry so far is illustrated in Fig. \ref{trproofcase2}.

It is easily verified that $\|x_1 - u\| \leq \|x_0 - u\|$ and $\|x_1 - v\| \leq \|x_0 - v\|$. Also, since the angle $\widehat{x_0 x_1 w}$ is at least  $90^{\circ}$, we have $\|x_1 - w\| \leq \|x_0 - w\|$, and therefore, $x_1 \ll x_0$. If $x_1 \in \mathcal{T}$, the lemma then holds for $y=x_1$. Otherwise, we consider the following three cases: The first case $w_1 \leq v_1$ is the same scenario as illustrated in Fig. \ref{trproofcase2}. In this case, we have $v \ll x_1$, and since $v\in \mathcal{T}$ obviously, the proof is complete with $y = v$. The second case $v_1 \leq w_1 \leq x_{01}$ is illustrated in Fig. \ref{trproofcase1}. We let $x_2 = \left[\begin{smallmatrix} w_1 \\ 0 \end{smallmatrix}\right]$, and $x_3$ to be the projection of $x_2$ on the edge $vw$.

The relation $x_2 \ll x_1$ obviously holds. The relation $x_3 \ll x_2$ follows from the same arguments that we have used to prove $x_1 \ll x_0$ in Fig. \ref{trproofcase2}. Since $x_3\in \mathcal{T}$, the lemma follows with $y = x_3$. Finally, for the third case $w_1 \geq x_{01}$, let $x_4$ to be the projection of $x_1$ on the edge $vw$. The proof of the relation $x_4 \ll x_1$ similarly follows the proof of $x_1 \ll x_0$ in Fig. \ref{trproofcase2}. The lemma then holds for $y = x_4$. This concludes the proof.

\begin{center}
\textsc{Appendix C: Proof of Proposition \ref{exactsolutionprop}}\vspace{-5pt}
\end{center}

Let $\phi_1(x) = |x-u| + |x - v|$ and $\phi_2(x) = c |x-w|^2$. Without loss of generality, let $u \leq v$. We have $\phi_1(x) \geq v-u$ with equality if and only if $x\in[u,v]$, and $\phi_2(x) \geq 0$ with equality if and only if $x = w$. Therefore, $\phi(x) \geq v-u$ with equality if and only if $x = w$ and $x\in[u,v]$, or equivalently, if $x = w$ and $w\in[u,v]$. This proves the first case in (\ref{xstars}). 

Suppose $w > v$. We first show that $x^{\star} \in [v, w]$. We have $x^{\star} \in [u,w]$ by Lemma \ref{trianglelemma}. Moreover, for any $x\in[u,v]$, we have $\phi(x) = v-u + c|x-w|^2 \geq v-u + c|v-w|^2$ with equality if and only if $x = v$. Hence, $x^{\star}\in[u,v]$ implies $x^{\star} = v$. Combining with $x^{\star} \in [u,w]$ yields $x^{\star} \in [v, w]$.

Now, let $\xi(x) = |2x-u-v| + c|x-w|^2$. We have $\xi(x) \leq \phi(x)$ for all $x \in \mathbb{R}$. Equality holds if and only if $x \leq u$ or $x \geq v$. Let $y^{\star} = \arg\min_{x\in\mathbb{R}} \xi(x)$ denote the global minimizer of $\xi$. According to \cite{proximal1}, we have $y^{\star} = (w - \alpha) \in [u,w]$. If further $y^{\star} \in [v,w]$, we have $y^{\star} = \arg\min_{x\in[v,w]} \xi(x) = \arg\min_{x\in [v,w]}\phi(x)  = \arg\min_{x\in\mathbb{R}}\phi(x) = x^{\star}$. Otherwise, if $y^{\star} \leq v$,  first note that $\xi$ is increasing on $[v,\infty)$ as $\xi$ is convex. It follows that $\phi$ is increasing on $[v,\infty)$. Since $x^{\star} \in [v, w]$ as already shown,  $\phi$ attains its minimum at $x^{\star} = v$. Hence, we have $x^{\star} = \max\{v,y^{\star}\} = \max\left\{v', w - \alpha \right\}$ in general, and this proves the second case in (\ref{xstars}). The final case in (\ref{xstars}) follows from the same arguments.

\begin{IEEEbiography}[{\includegraphics[width=1in,height=1.25in,clip,keepaspectratio]{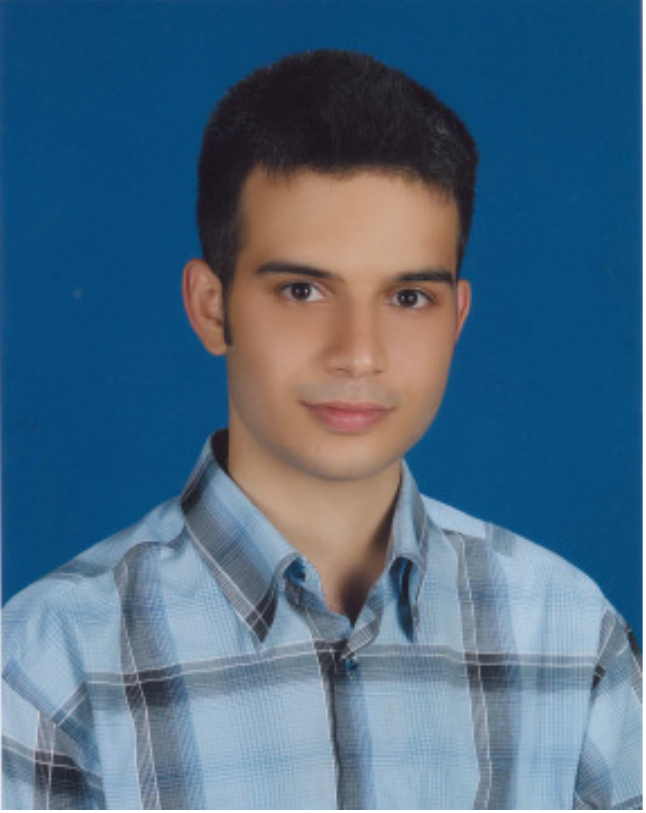}}]{Erdem Koyuncu} is an Assistant Professor at the Department of Electrical and Computer Engineering (ECE) of the University of Illinois at Chicago (UIC). He received the B.S. degree from the Department of Electrical and Electronics Engineering of Bilkent University in 2005. He received the M.S. and Ph.D. degrees in 2006 and 2010, respectively, both from the Department of Electrical Engineering and Computer Science of the University of California, Irvine (UCI). Between Jan. 2011 and Aug. 2016, he was a Postdoctoral Scholar at the Center for Pervasive Communications and Computing of UCI. Between Aug. 2016 and Aug. 2018, he was a Research Assistant Professor at the ECE Department of UIC. His research interests are in the areas of communications and signal processing.
\end{IEEEbiography}

\begin{IEEEbiography}[{\includegraphics[width=1in,height=1.25in,clip,keepaspectratio]{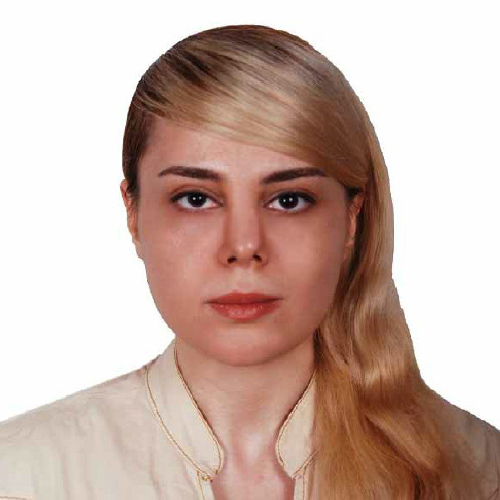}}]{Maryam Shabanighazikelayeh} received her B.Sc. in Electrical Engineering from K. N. Toosi University of Technology, Tehran, Iran in 2010 and her M.S. in Electrical Engineering from Sharif University of Technology, Tehran, Iran in 2013. She also worked as an RF optimization engineer in Huawei Technologies. She is currently a PhD candidate in Electrical Engineering, University of Illinois at Chicago. Her research interests include wireless communications, reinforcement learning and deep learning.
\end{IEEEbiography}

\begin{IEEEbiography}[{\includegraphics[width=1in,height=1.25in,clip,keepaspectratio]{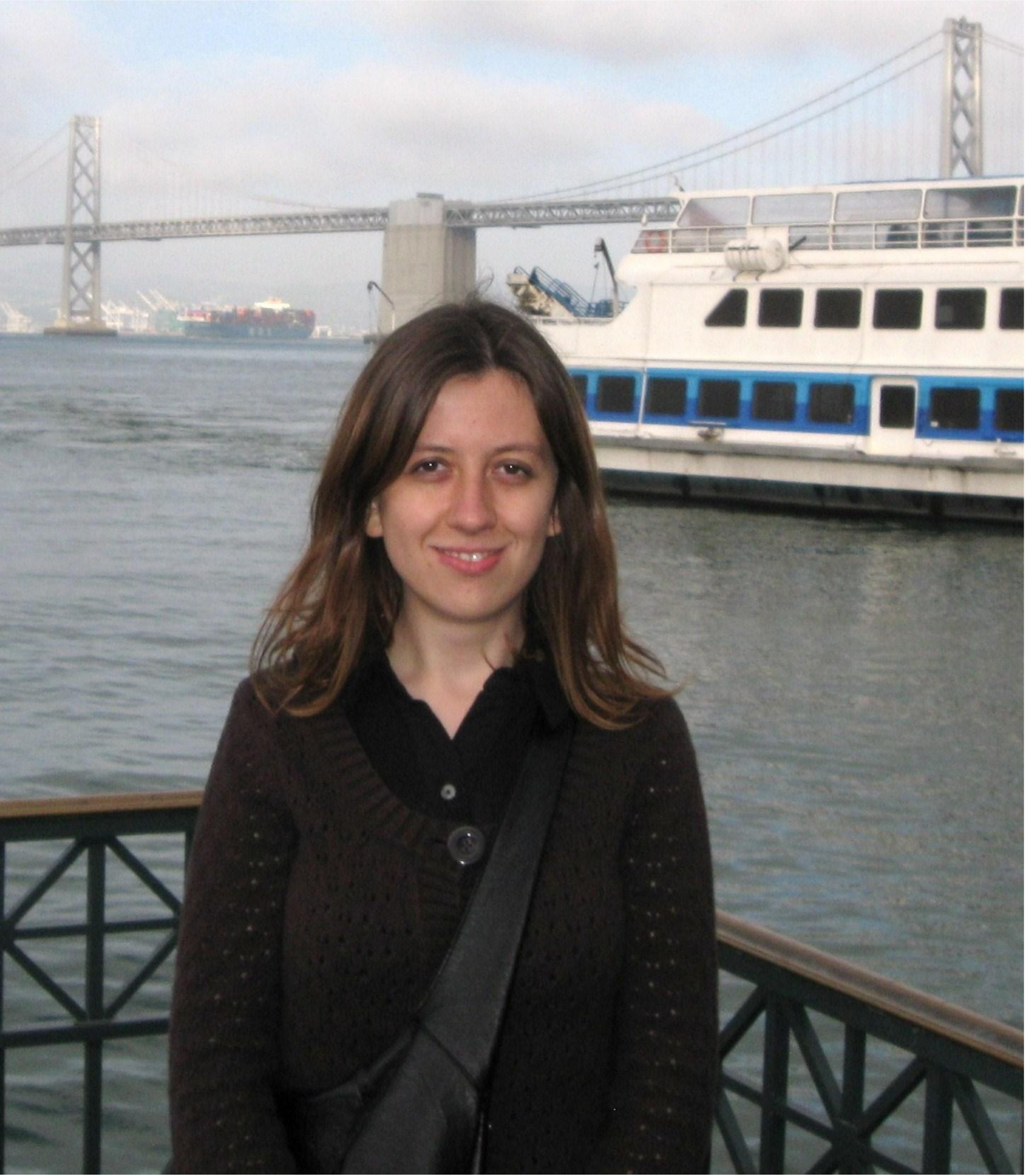}}]{Hulya Seferoglu} is an Assistant Professor in the Electrical and Computer Engineering Department of University of Illinois at Chicago. She received the B.S. degree in Electrical Engineering from Istanbul University, Turkey, in 2003, M.S. degree in Electrical Engineering and Computer Science from Sabanci University, Turkey in 2005, and Ph.D. degree in Electrical and Computer Engineering from University of California, Irvine in 2010. She worked as a Postdoctoral Associate in the Laboratory of Information and Decision Systems (LIDS) at Massachusetts Institute of Technology during 2011-2013. She worked as a summer intern at AT\&T Labs Research, Docomo USA Labs, and Microsoft Research Cambridge in 2010, 2008, and 2007, respectively. Her research interests are in the area of networking: design, analysis, and optimization of network protocols and algorithms. She is particularly interested in edge computing, network optimization, network coding, and multimedia streaming.
\end{IEEEbiography}

\end{document}